\documentclass[11pt,a4paper]{article}

\usepackage{jheppub}
\usepackage[hyperindex,breaklinks]{hyperref} 
\usepackage{graphicx}% Include figure files
\usepackage{dcolumn}% Align table columns on decimal point
\usepackage{bm}% bold math

\usepackage{atlasphysics}
\usepackage{url}
\usepackage{booktabs}
\usepackage{multirow}
\usepackage{tabularx}
\usepackage{subfigure}
\usepackage{lscape}
\usepackage{longtable}
\usepackage{rotating}

\RequirePackage{xspace}
\usepackage{relsize}

\newcommand{\Of}{\Omega^4}
\newcommand{\So}{\Sigma}
\newcommand{\bmu}{\mathcal{B}_\mu}
\newcommand{\gmu}{\mathcal{G}_\mu}
\newcommand{\wmu}{\mathcal{W}_\mu}
\newcommand{\hmu}{\mathcal{H}_\mu}
\newcommand{\qmu}{\mathcal{Q}_\mu^5}
\newcommand{\ymu}{\mathcal{Y}_\mu^5}

\newcommand{\oh}{\textstyle \frac{1}{2}}
\newcommand{\gM}{\gamma^\mu}
\newcommand{\gm}{\gamma_\mu}

\def\ipb{\mbox{pb$^{-1}$}}%  Inverse picobarns.

\begin{document}
%\pagewiselinenumbers
%\linenumbers
\bibliographystyle{JHEP}

{\flushright{\normalsize{CERN-PH-EP-2012-020}}}
\title{Search for same-sign top-quark production and fourth-generation down-type quarks in $pp$ collisions at $\sqrt{s}=7$~\tev\ with the ATLAS detector}

\abstract{
  A search is presented for same-sign top-quark production and
  down-type heavy quarks of charge $-1/3$  in events with two isolated leptons
  ($e$ or $\mu$) that have the same electric charge, at least two jets
  and large missing transverse momentum. The data are
  selected from  $pp$  collisions at $\sqrt{s}=7$~\tev\  recorded by the ATLAS detector and correspond
  to an integrated luminosity of 1.04~\ifb.  The observed data are
  consistent with expectations from Standard Model processes.
 Upper limits are set at 95\% confidence level on the cross section
  of new sources of
 same-sign top-quark pair production of 1.4-2.0 pb depending on the
 assumed mediator mass.  Upper limits are also set on the pair-production cross-section for new heavy down-type quarks; a lower limit of 450 \gev\ is set at 95\%
 confidence level on the mass of heavy down-type quarks under the
 assumption that they decay 100\% of the time to $Wt$.
}

\maketitle

\date{\today}

\section{Introduction}

The Standard Model (SM) of the electroweak and strong interactions is
extremely successful in explaining most of the measurements in
particle physics at energies accessible today. Its predicted behaviour
at high energies, however, presents some theoretical problems which
have motivated a large variety of theories encompassing and extending
it. Due to the large variety of models proposed, signature-based searches
are often useful when exploring the consequences of these theories in
an economical way. In hadron collisions it is useful to group final
states by the number of charged leptons (electrons or muons). Within
this classification, a signal with two leptons of the same electric
charge (same-sign leptons) is interesting since it has a low
background rate in the Standard Model, and potentially large
contributions from new theories, for example new flavour-changing $Z′$
bosons, proposed~\cite{Jung:2009jz} to explain the forward-backward
asymmetry ($A_{\textrm{FB}}$) measured at the Tevatron
~\cite{Aaltonen:2011kc,Collaboration:2011rq}, or new heavy quarks~\cite{Contino:2008hi,t53}.

\begin{figure}
\begin{center}
\begin{tabular}{ccccc}
\raisebox{-1.2cm}{\epsfig{file=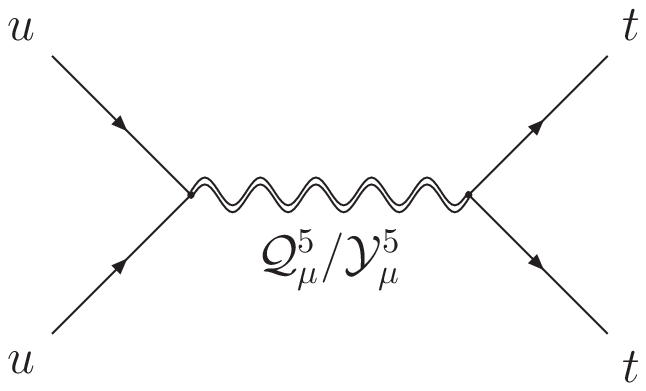,height=2.4cm}} &
$\stackrel{M \gg v}{\longrightarrow}$ &
\raisebox{-1.2cm}{\epsfig{file=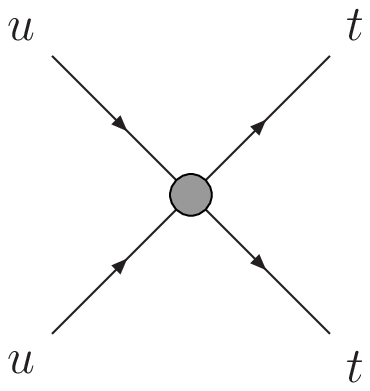,height=2.4cm}} &
$\stackrel{M \gg v}{\longleftarrow}$ &
\raisebox{-1cm}{\epsfig{file=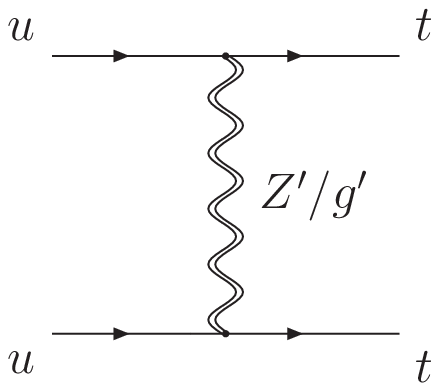,height=2.4cm}}
\end{tabular}
\end{center}
\caption{Production of same-sign top-quark pairs via the production of a heavy
  vector boson (such as color-triplet $\mathcal{Q}_\mu^5$ or
  color-sextet $\mathcal{Y}_\mu^5$~\cite{sseffop}) in the
  $s$-channel (left) or exchange of a heavy vector boson (such as $Z'$ or
  $g'$) in the $t$-channel (right) . For large resonance masses, both cases can be described by a four-fermion interaction (middle).}
\label{graph}
\end{figure}

\begin{figure}
\begin{center}
  \includegraphics[width=0.4\linewidth]{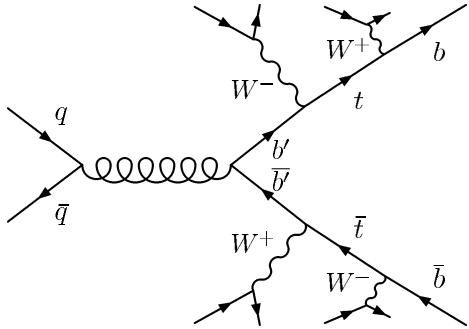}
\end{center}
\caption{ Pair production and decay of heavy quarks with decays to
  $W^+W^-\bar{b}W^-W^+b$.}
\label{graph2}
\end{figure}
In this paper we present a search for events characterised by two isolated same-sign leptons in association with at
least two jets and large  missing transverse momentum (\met). Two specific signal processes are considered, same-sign top-quark production \cite{Berger:2010fy,sstops} and pair production of 
 down-type heavy quarks of charge $-1/3$~\cite{Frampton:1999xi}. Feynman diagrams of these processes are
 shown in figures \ref{graph} and  \ref{graph2}, respectively.  
 The $uu \to tt$ process illustrated in figure 1 can be mediated at
 the tree level by the exchange of a $s$-channel resonance (left), or
 a $t$-channel resonance (right). In the case of new vector bosons exchanged in the $s$-channel, the new 
particle must be a colour-triplet or colour-sextet (respectively labelled as 
$\mathcal{Q}_\mu^5$, $\mathcal{Y}_\mu^5$) with charge $4/3$, while for $t$-channel exchange it can be a colour-singlet $Z'$ or colour-octet $g'$, both with zero 
charge. For resonance masses $m$ much
larger than the electroweak symmetry breaking scale $v$ and the typical energy scales in the process, all these cases can be described by a gauge-invariant effective 
four-fermion interaction, as shown in figure 1 (middle). For the heavy
quark search, a specific model 
 in which the heavy quark is a fourth-generation chiral quark is taken
 as representative and referred to as $b'$.  The search uses data recorded  by the ATLAS detector
from $pp$ collisions at a centre-of-mass energy of $\sqrt{s} = 7$
\tev\ produced by the Large Hadron Collider (LHC) with an integrated
luminosity of $1.04 \pm 0.04$~\ifb~\cite{lumi}.

The CMS and CDF Collaborations searched for fourth-generation
down-type quarks with same-sign leptons using 34 \ipb\ of $pp$ collisions~\cite{cms:2011em}
and 2.7\,\ifb\ of $p\bar{p}$ collisions~\cite{Aaltonen:2009nr}, respectively.
They set lower mass limits of 361~\gev\ and 338~\gev, respectively, at
95\% confidence level. The ATLAS and CDF Collaborations searched for
fourth-generation down-type quarks in single-lepton events with many
jets using 1.1 \ifb\ of $pp$ collisions~\cite{atlasljets} and 4.8 \ifb\
of $p\bar{p}$ collisions~\cite{cdfljets}, respectively. They set lower mass limits of 480~\gev\ and 372~\gev, respectively, at
95\% confidence level.

In this analysis, the  data are found to
be consistent with SM expectations, and upper limits on the same-sign
top quark production cross section are presented. These limits are
interpreted as constraints on the coefficients for a set of dimension-six
effective operators~\cite{sstops2} that can be used to parameterise same-sign top-quark production as four-fermion contact interactions. Limits on these coefficients are
translated into limits on a wide range of SM extensions mediating
same-sign top-quark production at the tree level~\cite{sseffop},
assuming that the new particles are heavy, which is consistent with
the non-observation of an excess over the SM prediction. Limits are
also obtained for the specific case of light
flavour-changing $Z'$ bosons~\cite{Jung:2009jz}.  Additionally,
upper limits are placed on the cross section of pair production of
$b'$, and a lower limit on the heavy quark mass is presented.

\section{The ATLAS detector}

The ATLAS detector~\cite{atlasdet} is a multipurpose detector with precision tracking,
calorimetry and muon spectrometry. The transverse momenta (\pT) of charged particles with
pseudorapidity $|\eta| < 2.5$\footnote{ATLAS uses a right-handed system with its origin at the nominal interaction point (IP) in the center of the detector and the $z$-axis along the beam pipe. The $x$-axis points from the IP to the center of the LHC ring, and the $y$-axis points upward. Cylindrical coordinates ($r$,$\phi$) are used in the transverse plane, $\phi$ being the azimuthal angle around the beam pipe. The pseudorapidity is defined in terms of the polar angle $\theta$ as $\eta  = -\ln \tan(\theta/2)$. Distances in $\eta-\phi$ space are given as $\Delta R \, = \, \sqrt{\Delta\phi^2 + \Delta\eta^2}$.} are measured by the inner detector (ID),
which is a combination of a silicon pixel detector, a silicon
microstrip detector and a straw-tube detector. The ID operates in a
uniform 2 tesla magnetic field. Measurements from the pixel detector   enable precise determination of production vertices.
Electromagnetic calorimetry for electron, photon, and jet
reconstruction is provided by a high granularity, three layer depth-sampling
liquid-argon (LAr) detector with lead absorbers in the region $|\eta|
< $ 3.2. Jet reconstruction also uses hadron calorimetry provided by a
scintillating tile detector with iron absorbers in the central region
for $|\eta| < $ 1.7, and a LAr  active-medium sampling calorimeter for 1.5 $< |\eta| < 4.9$. A presampler detector is used to correct for energy losses by electrons and photons in material in front of the calorimeter for $|\eta| < 1.8$.

Muons are detected with a multi-system muon spectrometer
(MS). Precision measurements of the track coordinates are provided by
monitored drift tubes over most of the $\eta$ range. These are
supplemented by cathode-strip chambers measuring both the $\eta$ and
$\phi$ coordinates for $2.0 < |\eta| < 2.7$ in the innermost
endcap muon station. Fast measurements required for initiating trigger
logic are provided by resistive-plate chambers for $|\eta|< 1.05$, and
beyond that by thin-gap chambers for $|\eta| < 2.4$. The muon
detectors operate in a non-uniform magnetic field generated by three
superconducting air-core toroid magnetic systems with eight coils per toroid.

To trigger readout, full event reconstruction and event storage by the
data acquisition system, electron candidates must have transverse
energy greater than 20 GeV. They must satisfy shower-shape
requirements and correspond to an ID track. Muon candidates must have
transverse momentum greater than 18 GeV and a consistent trajectory
reconstructed in the ID and muon spectrometer. The full trigger chain
uses signals from all muon detectors. These triggers reach their
efficiency plateau at lepton \pT\ thresholds of 20 GeV for muons and
25 GeV for electrons.

\section{Data and Monte Carlo Samples}

\subsection{Data Sample}
The data used in this search were collected by the ATLAS detector at
the CERN LHC between March and June of 2011, using a single muon or electron trigger as
described above.  The data sample corresponds to an integrated luminosity of
$1.04 \pm 0.04$ \ifb~\cite{lumi}.

\subsection{Monte Carlo Samples}

Monte Carlo simulation samples have been used to develop and validate
the analysis procedures, calculate the acceptance for signal events
and to evaluate the contributions from some background processes. The
ATLAS software~\cite{:2010wqa} uses {\sc
  geant}4~\cite{Ivanchenko:2003xp} to simulate the
detector response.

In order to describe properly the
effects of multiple proton-proton interactions per bunch crossing, the
Monte Carlo samples contain multiple interactions per beam-crossing,
weighted to match the data.  Except where specifically noted,
all simulated samples are generated with the CTEQ6L1~\cite{cteq}
parton distribution functions (PDF). Simulated samples of same-sign top-quark production with dileptonic
decay  have been generated by {\sc protos}~\cite{sstops2}, accurate
to leading order (LO) in QCD, with showering and hadronisation performed by {\sc
  pythia}~\cite{Sjostrand:2006za}.  Samples have been generated for
each of the three possible chirality configurations (left-left, left-right, right-right).

Simulated samples of heavy down-type quark pair production and decay  have been
generated 
by {\sc pythia} using the MRST2007 LO* PDF
set~\cite{Sherstnev:2007nd}  for several mass values, between 300 and
600 GeV; the cross section is normalized
to NNLO~\cite{hathor}. 

Several background processes contribute to the final state of
same-sign leptons with associated jets. The largest backgrounds
(including top-quark pair production, $W$+jets and single top quark
production) are
estimated from data, as described in detail below (thereafter
referred to as \mbox{`data-driven'}).
Additional background estimates are described using simulated Monte
Carlo samples as listed here:
 
\begin{itemize}

\item Di-boson production ($W^{\pm}W^{\mp}, WZ, ZZ$) was generated using {\sc alpgen}~\cite{alpgen}
  to explicitly account for hard emission  of up to two partons and {\sc
    herwig}~\cite{herwig} to describe soft emission, showering and
  hadronisation. The cross sections are normalised to next-to-leading-order (NLO) theoretical calculations~\cite{mcfm}.
\item $t\bar{t}+W$, $t\bar{t}+Z$, $t\bar{t}+W$+jet, $t\bar{t}+Z$+jet,
  $t\bar{t}+W^{\pm}W^{\mp}$, $W^{\pm}W^{\pm}+$2 jets were generated with
  {\sc madgraph}~\cite{Alwall:2007st}, and showered and hadronised with
  {\sc pythia}. These are normalised to LO theoretical calculations~\cite{Alwall:2007st}.
\end{itemize}

\section{Object Reconstruction}

Electrons are found by a calorimeter-seeded reconstruction algorithm
and are matched to a track.  They are
required to satisfy $E_{\textrm{T}} =
E_{\textrm{cluster}}$/cosh($\eta_{\textrm{track}}) >$  25 GeV (where
`cluster' refers to the calorimeter electron cluster) in a
pseudorapidity range $|\eta_{\textrm{cluster}}| < $ 2.47 but excluding
the transition region between the barrel and endcap calorimeters
covering 1.37 $< |\eta_{\textrm{cluster}}| <$ 1.52.  We require a `tight' electron
selection~\cite{egamma}. Electrons must also satisfy calorimeter isolation: the difference
between the transverse energy deposited inside a cone of size $\Delta R =
\sqrt{\Delta\phi^2+\Delta\eta^2}= 0.2$ around the electron direction
and the electron transverse energy has to be lower than \mbox{3.5 GeV}.

Muons are found with an algorithm which requires that tracks reconstructed
in the muon spectrometer match a track in the inner detector~\cite{staco}.  We apply
a loose cosmic ray rejection by removing all back-to-back muon pairs ($\Delta\phi(\mu_1,\mu_2) >$  3.1)
whose transverse
impact parameter with respect to the beam spot is greater than 0.5
mm. Muon candidates must satisfy \mbox{$p_{T}> 20$ GeV} and
\mbox{$|\eta| < 2.5$}.
We also require that the muons are isolated. An $\eta$-$\phi$ cone of
$\Delta R=0.3$ about the muon direction must contain less than 4 GeV
of additional energy in the calorimeter and less than 4 GeV from additional tracks. Finally,
we remove all muons within an $\eta$-$\phi$ cone of \mbox{$\Delta R$  = 0.4} of any jet with \mbox{\eT\ $>$ 20 GeV}.

Jets~\cite{atlasjets} are reconstructed from topological clusters of calorimeter energy deposits~\cite{topocluster} using the anti-$k_t$
algorithm~\cite{antiktalgorithm} with a radius parameter equal to
0.4.  A jet energy scale (JES) correction is applied to account for the energy response and non-uniformity of the EM and hadronic calorimeters. Jets are required to satisfy \pT\ $>$ 20 GeV and $|\eta| < 2.5$. Jets overlapping with selected electrons
within an $\eta$-$\phi$ cone of $\Delta R$ = 0.2 are removed.  During
part of the data-taking period, data from a small portion of the LAr electromagnetic
calorimeter was not read out due to a technical problem; this results in a signal
efficiency loss of 10-15\%, depending on jet multiplicity.

Missing transverse momentum (\MET) is constructed from
the vector sum of topological \mbox{calorimeter} cluster deposits, projected in the
transverse plane~\cite{met}.  Deposits associated with
selected jets or electrons are corrected to the energy scale
appropriate for jets or electrons, respectively. Muons are included in
the \MET\ calculation after a correction for the
muon contribution to calorimeter energy
deposits. All other energy deposits with $|\eta|<4.5$ and not
associated with leptons or jets contribute to the calculation of \MET.

\section{Event Selection}
\label{sec:selec}

In the mass range considered, where $m_{b'}>m_t+m_W$, each $b'$ is
assumed to
decay exclusively to a top quark and a $W$ boson, giving a signature of
$t\bar{t}$ with two additional $W$ bosons, as shown in
figure~\ref{graph2}.  The signature for same-sign top-quark production is
similar to $t\bar{t}$, but with two positive leptons in the final
state from top-quark decay due to the asymmetric charge of the $pp$
 collisions. In this analysis, the final state for both signatures must contain at least
one lepton pair with the same electric charge, plus missing transverse
momentum from neutrinos, and a large jet multiplicity.

Events are selected that satisfy the following requirements:

\begin{itemize}
\item  Events must contain a primary vertex,  consistent with the beam spot position,
determined with at least five tracks, each with \pT\ $>0.4$ GeV;
\item Events must contain at least two leptons with the same electric
  charge, each within $|\eta|<2.5$. Muons must have \pT\ $>20$ GeV
  and electrons must have \eT\ $>$  25 GeV. In events with more than one
  same-sign pair, the pairs are sorted according to the leading lepton
  \pT, then by the subleading lepton \pT. The first pair is chosen;
\item In the $ee$ or $\mu\mu$ channel, the invariant mass of the two
  leptons must exceed 15 GeV and not be in the $Z$-boson mass window: $|m_{\ell\ell}-m_{Z}|>10$~GeV;
\item Events must contain at least two jets, each with \pT\ $>$ 20 GeV and $|\eta|<2.5$;
\item The magnitude of the missing transverse momentum must be greater than 40 GeV;
\item Three signal regions are defined, as follows.
\begin{itemize}
\item For a heavy down-type quark, and for same-sign top quarks
  produced from high-mass $Z'$ exchange,  the scalar sum \HT\ of the
  transverse energy of the selected
leptons and jets must exceed 350 GeV. This cut has been optimised in
  order to reach the maximum sensitivity with $m_{b'}= 400$ GeV (close
  to previous exclusion limits) and  for same-sign top quarks
  produced from high-mass $Z'$ exchange. Including both lepton
  charge configurations, we refer to this as the
  `heavy-quark signal region'.
\item When applied to searches for same-sign top-quarks, the events
  are required to satisfy all the requirements of the heavy-quark
  signal region, but including only events with positively-charged leptons, as the $pp$
  initial state of the LHC gives  predominantly positively-charged top
  quarks; we refer to this as the `same-sign top-quark signal region'.
\item For same-sign top quarks produced from low-mass $Z'$ exchange, the
  signal region is optimised by requiring positively-charged leptons, \HT\ $>$ 150 GeV and invariant mass of
  the lepton pair $m_{\ell\ell}>$  100~GeV; we refer to this as the
  `low-mass $Z'$ boson signal region'.
\end{itemize}
\end{itemize}

The efficiencies of the event selection for heavy-quark  and
same-sign top-quark events are given in Table~\ref{tab:eff}.

\begin{table}
\center
\begin{tabular}{lcc} \hline\hline

\multicolumn{3}{c}{ Heavy-quark and same-sign top-quark signal regions}\\\hline
& \multicolumn{2}{c}{$\ell^+\ell^+$ and $\ell^-\ell^-$}\\ 
$b'_{m_{b'}=350\ {\rm GeV}}$ & \multicolumn{2}{c}{$2.0\%$} \\
$b'_{m_{b'}=450\ {\rm GeV}}$ & \multicolumn{2}{c}{$2.5\%$} \\
$b'_{m_{b'}=550\ {\rm GeV}}$ & \multicolumn{2}{c}{$2.7\%$} \\\cline{2-3}
 & $\ell^+\ell^+$ & $\ell^-\ell^-$ \\ 

$tt_{LL}$ & $0.7\%$ & negligible \\
$tt_{LR}$ & $0.8\%$ & negligible \\
$tt_{RR}$ & $0.8\%$ & negligible \\ \hline
\multicolumn{3}{c}{ Low-mass $Z'$ boson signal region}\\ \hline
$tt_{RR,m_{Z'}=100\ {\rm GeV}}$ & $0.7\%$& negligible \\
$tt_{RR,m_{Z'}=150\ {\rm GeV}}$ & $0.8\%$& negligible\\
$tt_{RR,m_{Z'}=200\ {\rm GeV}}$ & $1.0\%$& negligible \\
\hline

\hline\hline
\end{tabular}
\caption{ Efficiencies of the event selection for heavy-quark  and
same-sign top-quark events in the heavy-quark signal region (two
same-sign leptons, at least two jets, and \MET\ $>$ 40~GeV and
$H_{\textrm T}>350$ GeV) as well as efficiencies of the event selection
for same-sign top-quark events via a low-mass $Z'$ boson in the
low-mass $Z'$ boson signal region (two
same-sign leptons, at least two jets, \MET\ $>$ 40~GeV,  $H_{\textrm
  T}>150$ GeV and $m_{\ell\ell}>100$ GeV). Efficiencies are relative
to the total cross section, and so include the effect of branching
ratios (4.5\% for $tt$ and 9\% for $b'$, without including
$W\rightarrow\tau\nu$ contributions) as well as
acceptance. Statistical uncertainty is $0.01\%$ for $tt$ and $0.05$\% for $b'$.}
\label{tab:eff}
\end{table}

\section{Standard Model Backgrounds}

In the SM, events with the same-sign dilepton signature are due to
three categories of processes: 
\begin{itemize}
\item those in which one lepton originates from a jet or from a photon conversion,
\item those with an opposite-sign
dilepton pair in which the reconstructed charge of one lepton is
mismeasured, and
\item those that originate from a pair of $Z/W$ gauge bosons.
\end{itemize}
\noindent
The diboson contribution is estimated using simulated
samples, and the remaining backgrounds are
estimated by extrapolation from control samples selected in the data
as described in the following sections.

\subsection{Backgrounds with leptons originating from jets or photons}

A significant SM background source is due to events in which one of the two leptons comes
from the decay of a $W$ or $Z$ boson (called `real' below) and the second is a `fake'
lepton, a jet or photon misreconstructed as an isolated lepton.  Here `fake' is
used to indicate both non-prompt leptons and misidentified $\pi^0$s,
conversions, etc. 

The dominant fake-lepton mechanism is the semi-leptonic decay of a
$b$- or $c$-hadron, in which a muon survives the isolation
requirements. In the case of electrons, the three mechanisms are $b$-
or $c$- hadron decay, light flavour jets with a leading $\pi^0$ overlapping
with a charged particle, and conversion of photons.  Processes that contribute are opposite-sign top-quark pair production, production of $W$ bosons in association with
jets and multi-jet production.

The `matrix method'~\cite{top2010} is applied to estimate the fraction
of events in the signal regions that contains at least one fake lepton. A
selection is defined to isolate lepton-like jets and used to count the number of
observed dilepton events with zero, one or two selected leptons (`L')
together with two, one or zero lepton-like jets (`J'), respectively
($N_{JJ},N_{LJ},N_{JL},N_{LL}$).   The categories $N_{LJ}$ and
$N_{JL}$ are distinguished by \pT-ordering.

Two probabilities are defined and measured: $r$ and $f$, the probabilities that
real or fake leptons, respectively, which satisfy the lepton-like jet selection also
satisfy the final lepton selection requirements. Using $r$ and $f$,
linear expressions are  obtained for the observed yields as a
function of the number of events with zero, one and two real leptons (`R')
together with two, one and zero fake leptons (`F'), respectively
($N_{FF},N_{RF},N_{FR},N_{RR}$, respectively). These linear expressions form a matrix that is inverted in order to
extract the real and fake content of the selected dilepton event
sample.   The categories $N_{RF}$ and
$N_{FR}$ are distinguished by \pT-ordering.

For muons, lepton-like jets are found by removing the three isolation
requirements: calorimeter, track and jet isolation as described above. For electrons, lepton-like jets are found by removing
the requirement on the electron isolation and the quality of the
associated ID track.

The probability for real leptons $r$ is measured in samples of opposite-sign dielectron and
dimuon events, with one selected lepton and one lepton-like jet,
which are dominated by $Z\rightarrow \ell\ell$ decays. The requirement
86 \gev\ $< m_{\ell^+\ell^-}<$ 96 \gev\ is applied to achieve a high purity.

  The corresponding probability for fake leptons $f$ is measured in data
from a sample of single-lepton candidate events dominated by multi-jet
production, where contributions from real leptons  are suppressed using kinematic requirements. The
probability $f$ is found to decrease significantly with rising lepton
$p_{\textrm T}$ or event $H_{\textrm T}$, and is therefore parameterised in these
variables~\cite{atlas2010ss}.  

   The value of the probability $f$ is found to be substantially
   different in samples with and without a large heavy-flavour
   contribution. If the heavy-flavour fraction in the signal
region is different from the control region, using the 
value of $f$ from the control region would lead to a biased estimate of the fake-lepton
contribution in the signal regions.  Instead, individual jets are
assigned either a heavy-flavour probability ($f_{\mathrm{HF}}$) or a
light-flavour probability ($f_{\mathrm{LF}}$) based on a standard
heavy-flavour tagging algorithm which identifies jets with tracks
which have large impact parameter significance~\cite{jetprob}. This
algorithm correctly identifies 90\%
of heavy-flavour jets and misidentifies 50\% of light-flavour jets in $t\bar{t}$
simulated events.  The probability  $f_{\mathrm{HF}}$ 
is measured in samples of heavy-flavour jets, selected using the same
tagging algorithm with a stricter requirement.  The probability  $f_{\mathrm{LF}}$ is measured in
a sample of jets constructed by requiring small impact parameter
significance.

\subsection{Background from charge misidentification}

Events in which a pair of opposite-sign leptons are produced may be reconstructed as
a pair of same-sign leptons if a lepton charge is misidentified.  This is referred to as `charge flip' in the plots and the tables. 

For electrons, the dominant mechanism is hard bremsstrahlung, producing a photon which carries a large fraction of
the electron momentum and then converts asymmetrically, giving most of
its momentum to an electron of the opposite charge. Processes susceptible to this effect
are those with opposite-sign electrons, such as $Z$/$\gamma^* \rightarrow e^+e^-$ (including $Z$/$\gamma^*
\rightarrow \tau^+\tau^- \rightarrow e^+e^- + 4\nu$)
and $t\bar{t}$ dileptonic decays with at least one
electron.   For both electrons and muons, misidentification of the charge may occur due to
misreconstruction of the ID track. For muons, the requirement that the
measured charges in the ID and MS agree reduces this to negligible levels in the
range of $p_{\textrm T}$ found in the sample. 

The charge-flip rate for electrons is derived as a function of
electron $\eta$ from the rate of
same-sign and opposite-sign electron pairs in events with
$m_{\ell\ell} \in [81,101]$~GeV.  This rate is then applied to events with
an opposite-sign $ee$ or $e\mu$ pair to model the charge-flip
contribution to the same-sign sample.

A fraction of the charge-flip electron background is included in the
data-driven estimate of the fake lepton events described above.
However, studies of $Z$ events show that electron charge-flips are
well modeled using the weighted opposite-sign lepton pair events as
described here. To avoid double counting, the
overlap is removed from the fake-lepton background prediction.  The charge-flip overlap fraction
is measured by normalizing the  prediction to the observed
same-sign dilepton peak from $Z \to e^+ e^-$ processes in events with
$m_{\ell\ell} \in [81,101]$~GeV and found to be $(23\pm$3$)$\%.

\subsection{Backgrounds from processes with two electroweak bosons}

True same-sign dilepton events are produced from SM diboson processes
such as $WZ$ or $ZZ$ production. With a total cross section to
same-sign leptons equal to 0.7 pb, small with
respect to the expected $b'$ or same-sign top-quark production, SM diboson events are a rare but irreducible background to new physics sources, 
since events with more than two leptons are not excluded by the
selection.  This category includes events from the processes $t\bar{t}+W$, $t\bar{t}+Z$, $t\bar{t}+W$+jet, $t\bar{t}+Z$+jet,
  $t\bar{t}+W^{\pm}W^{\mp}$, and $W^{\pm}W^{\pm}+$2 jets, which
  together contribute 12-29\% of the diboson background. The contribution to the selected sample is estimated using
simulated events, as described above, and referred to as `real' in
plots and tables.

\subsection{Background Control Regions}

To validate the modeling of the SM backgrounds,  two control
regions are examined. Control regions are orthogonal to signal
regions and defined by selections which suppress possible signal contributions.

The first control region inverts the charge selection, requiring a
pair of opposite-sign leptons, at least two
reconstructed jets and missing transverse momentum greater than
\mbox{40 GeV}. This region validates the lepton efficiencies and the
modeling of the missing transverse momentum and $H_{\textrm T}$.
Figure~\ref{fig:valid_os} shows the observed and expected missing transverse momentum and
$H_{\textrm T}$ distributions, which are in good agreement within uncertainties.

The second control region is used to validate modeling of the same-sign
background sources, but in events with no reconstructed jets.
Figure~\ref{fig:valid_ss} shows the observed and expected missing transverse momentum and
$H_{\textrm T}$ distributions, which are in good agreement within uncertainties.

\begin{figure}
 \includegraphics[width=0.49\linewidth]{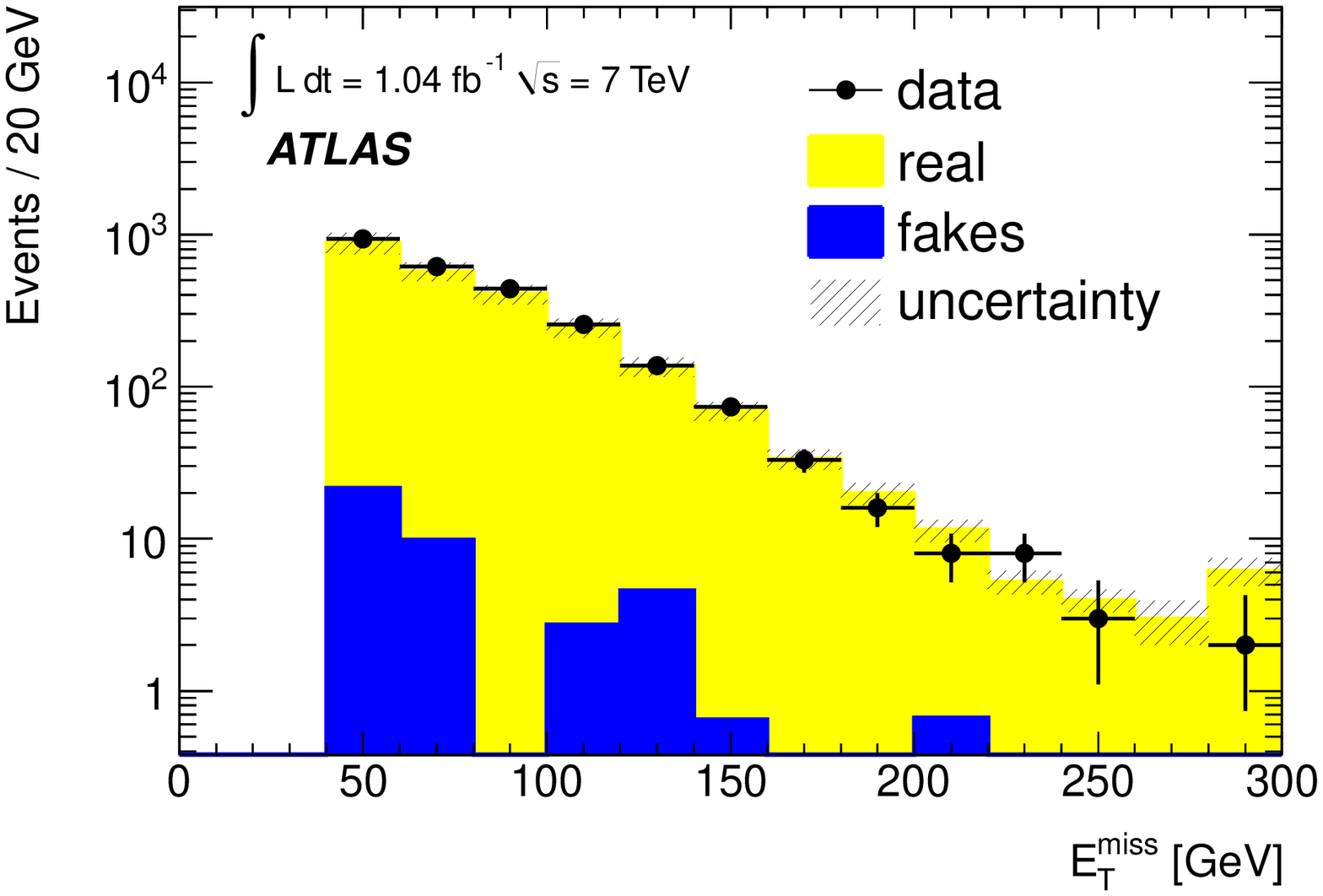}
  \includegraphics[width=0.49\linewidth]{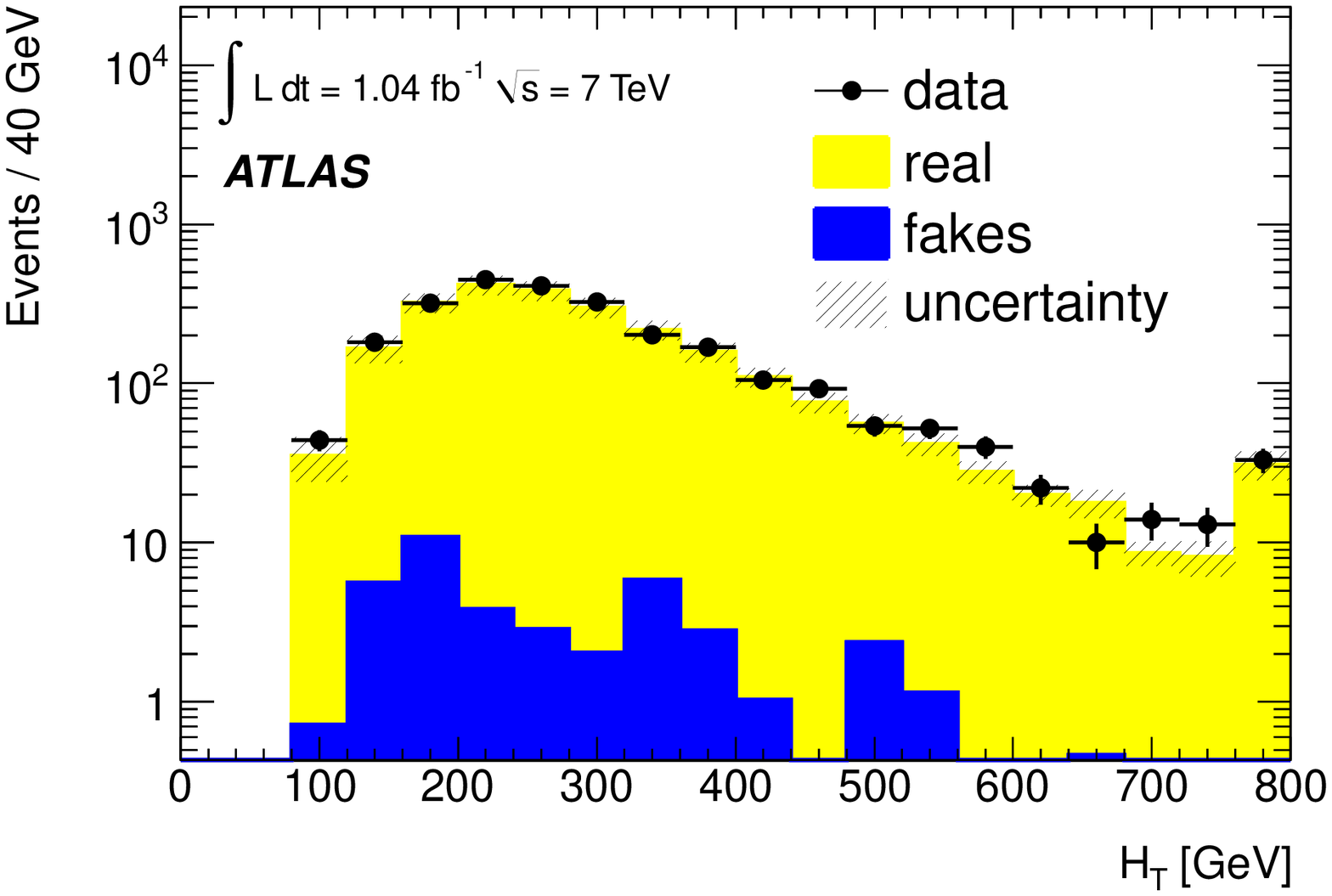}
\caption{ Comparison of observed data and expected SM backgrounds in
    events with a pair of opposite-sign leptons, at least two reconstructed
    jets and missing transverse momentum greater than 40
    GeV. Distribution of missing
    transverse momentum (left) and $H_{\textrm T}$ (right). Uncertainties (hatched) are systematic and statistical. The last bin
    includes overflow events. }
  \label{fig:valid_os}
\end{figure}

\begin{figure}
  \includegraphics[width=0.49\linewidth]{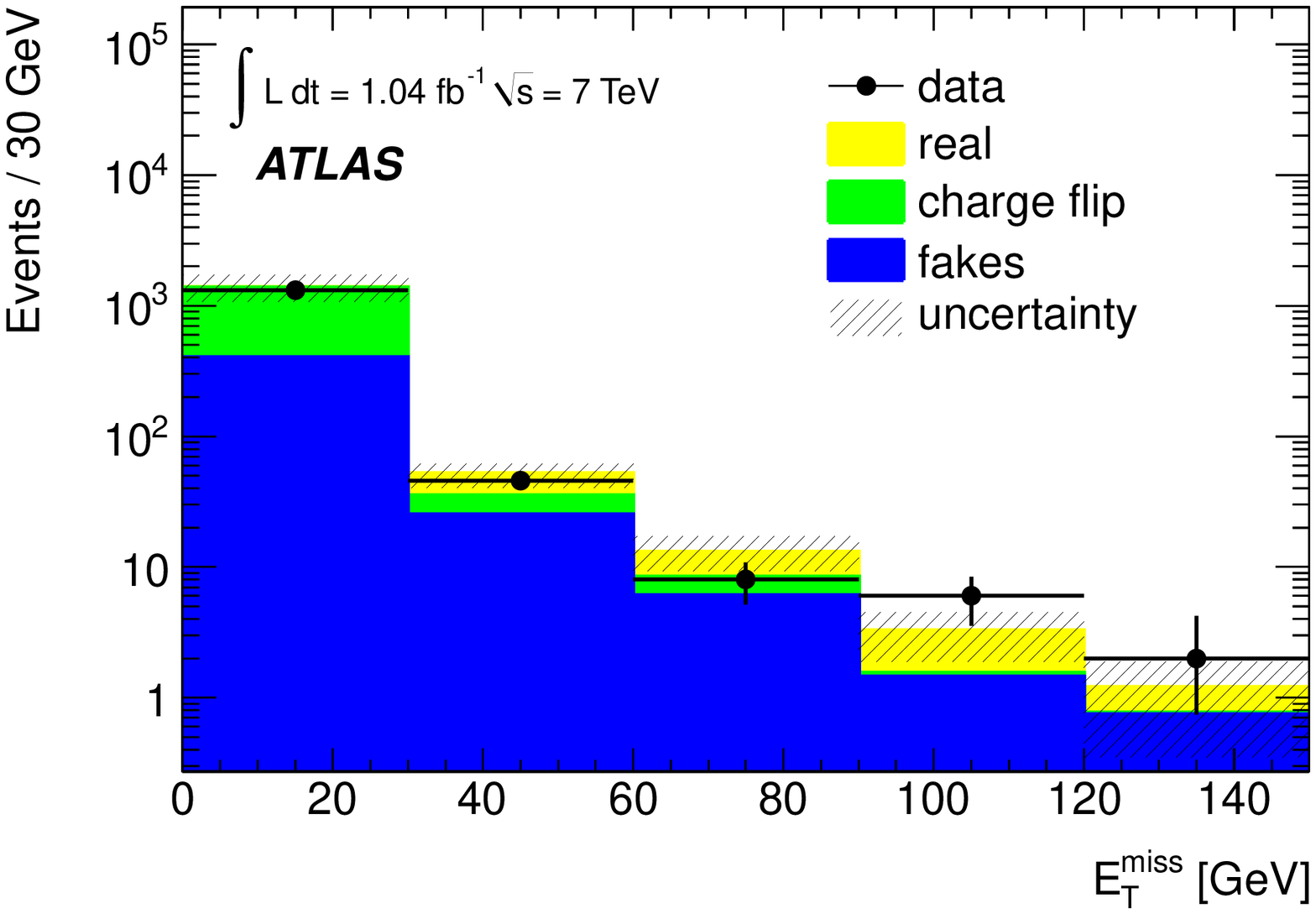}
  \includegraphics[width=0.49\linewidth]{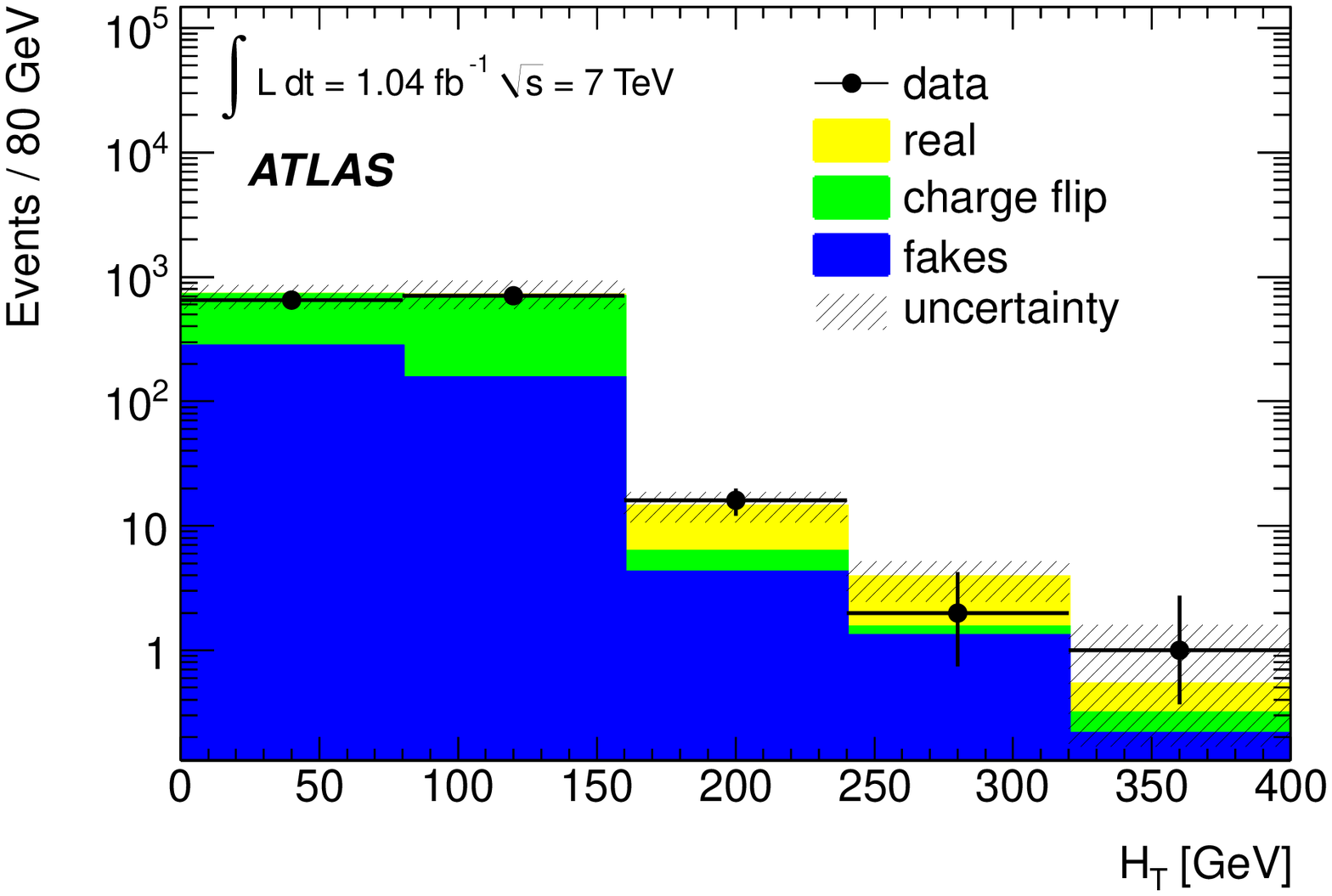}
  \caption{ Comparison of observed data and expected SM backgrounds in
    events with a pair of same-sign leptons and no reconstructed
    jets; the invariant mass requirement is not made here.  Distribution of missing transverse momentum (left) and $H_{\textrm T}$ (right). Uncertainties (hatched) are systematic and statistical. The last bin
    includes overflow events.}
  \label{fig:valid_ss}
\end{figure}

\section{Systematic uncertainties}
\label{sec:syst}

Several sources of systematic uncertainties have been considered, and their estimates are summarised below.
\begin{itemize}
\item Object calibration and resolutions:
Uncertainties in the jet~\cite{atlasjets} and lepton efficiency, energy or
momentum calibration, and resolution lead to
systematic uncertainties on the signal and background
acceptances.  There is also some uncertainty in
  estimating the effect of problems with the LAr calorimeter readout
  as described above. These uncertainties are summarised in
table~\ref{tab:syste} for signal and background separately.
\item Fake-lepton background: A 20\%-70\% uncertainty on the estimate
 of the fake lepton background is estimated from a combination of two
 sources. First, we vary the heavy-flavour tagging threshold defined in
 section~\ref{sec:selec} to modify
 the heavy-flavour identification efficiency  within its
 uncertainties; the difference in the estimated fake-lepton background is taken as an
 uncertainty.  Second, we compare the fake probability $f$ measured
 from the simulated multijet samples to probabilities measured in
 $t\bar{t}$ and $W$ boson+jets simulated samples; the difference in the
 probabilities is propagated to the fake-lepton background estimate.
\item Charge-flip background: the uncertainty on the overall scale of
  the charge-flip background in the signal regions is derived from a 
  comparison of the charge-flip rate extracted by several methods. 
All techniques use dielectron events with dilepton invariant mass
close to the $Z$ boson mass. The primary technique uses a maximum
likelihood fit to extract the charge-flip rates in different kinematic
regions simultaneously. An alternative method, tag-and-probe,
identifies a tag electron in the low pseudorapidity region which
satisfies strict track-matching requirements to ensure a negligible
charge-flip rate; the probe electron is used to measure the
charge-flip rate. In a third method, the rates for each region are
derived from electron pairs in that region.  The difference in
  the method results is used to estimate the systematic uncertainty, which is
  found to be $30\%-100\%$, increasing as a function of lepton $|\eta|$ and \pT.
\item Uncertainties affecting the Monte Carlo
  backgrounds and signals: luminosity and Monte Carlo cross
  sections. The uncertainty on the measured luminosity from van der
  Meer scans was estimated to be 3.7\%~\cite{lumi}. Uncertainties on
  Monte Carlo cross sections  depend on the process. The main source of systematic uncertainty on the production cross section originates from the diboson contribution.
   This uncertainty, which is nearly 100\%, is estimated from
   the difference between the nominal 
  sample described above and an alternative sample which uses {\sc herwig} to model the generation and emission of hard partons.
\item Initial and final state QCD radiation: parameters describing the
  level of radiation in the simulation are varied over a range
  consistent with experimental data~\cite{atlasperf}. In the signal regions, the
  corresponding uncertainty on the acceptance is $12.3\%$ for the $tt$
  signal and $6.8\%$ for the $b'$ signal.
\item Parton Distribution Functions: the uncertainty is evaluated using a
  range of current PDF sets~\cite{atlasperf}. In the signal regions, the
  uncertainty on the acceptance is $2.0\%$ for both the $tt$ and $b'$
  signals.

\end{itemize}
\begin{table}[htbp]
 \begin{center}
   \begin{tabular}{lccc}
     \hline     \hline
     Source  & Uncertainty on $tt$ & Uncertainty on $b'$  & Uncertainty on \\
             & acceptance (\%)      & acceptance (\%)          &
             background \\
             &       &           &  acceptance (\%) \\
      \hline
      Jet energy scale & ${+1.7}, {-2.3}$ & ${+0.5}, {-0.5}$ & ${+7.1}, {-5.5}$ \\ 
      Jet energy resolution& ${+0.7}, {-0.8}$  & ${+0.7}, {-0.8}$ & ${+4.1}, {-4.2}$ \\ 
      Jet reconstruction efficiency & $+0.1, -0.1$ & $+0.1, -0.1$ & $+0.1, -0.1$\\
      Electron energy scale  & ${+0.5}, {-0.7}$& ${+0.2}, {-0.3}$ & ${+1.1}, {-1.2}$ \\ 
      Electron energy resolution & ${+0.1}, {-0.3}$   & ${+0.1}, {-0.1}$ & ${+0.8}, {-0.8}$ \\
      Electron efficiency & ${+3.5}, {-4.2}$ & ${+3.3}, {-4.9}$ & ${+1.8}, {-2.0}$ \\ 
      Muon energy scale  & ${+0.1}, {-0.1}$ & ${+0.1}, {-0.1}$ & ${+0.1}, {-0.1}$ \\ 
      Muon energy resolution  & ${+0.2}, {-0.3}$ & ${+0.1}, {-0.2}$ & ${+0.6}, {-0.7} $\\ 
      Muon efficiency & ${+4.7}, {-5.8}$ & ${+4.3}, {-5.2}$ & ${+2.3}, {-3.0}$ \\ 
      Missing transverse momentum & ${+0.8}, {-0.8}$ & ${+0.8}, {-0.8}$ & ${+0.9}, {-0.9}$ \\ 
      LAr calorimeter readout & ${+2.0}, {-3.0}$ & ${+2.0}, {-3.0}$ & ${+1.0}, {-2.0}$\\
      \hline \hline
   \end{tabular}
 \caption{Sources of systematic uncertainties related to jet and lepton energy or momentum calibration and resolution,
and their contributions   to the uncertainty
    \mbox{on signal ($tt$ and $b'$) and background acceptance}. Lepton
   efficiencies  include trigger, reconstruction and identification terms.}
 \label{tab:syste}
 \end{center}
\end{table}

\clearpage
\section{Results}
\label{sec:consistency}

Due to the charge of the initial state in $pp$ collisions,
the SM and same-sign top-quark expectation for positively- and negatively-charged pairs are not
equal.  This makes the negatively-charged pairs a control region for
the same-sign top-quark signal. In both cases, the observed number of events agrees well with the SM
expectation within uncertainties.  Heavy down-type quarks are expected
in both the positively- and negatively-charged samples.

In each of the three signal regions, the largest source of SM background is due
to diboson production, followed by the fake lepton background.   In
table~\ref{tab:sig_neg} (table~\ref{tab:sig_pos}) the expected and observed yields are shown for
events with two negatively (positively) charged leptons in the
heavy-quark signal region. The positive-lepton sample is the same-sign
top-quark signal region. The
distributions of \MET\ and \HT\ are shown in figures~\ref{fig:sig_ss} and ~\ref{fig:sig_ss2}.

\begin{table}
\center
{\small
\begin{tabular}{lrrr}
\hline\hline
 & $e^-e^-$  & $\mu^-\mu^-$  & $e^-\mu^-$ \\ \hline
Fake
& $0.2 \pm 0.3  \pm 0.1 $ 
& $0.7 \pm 0.3  ^{+0.6}_{-0.3} $ 
& $0.5 \pm 0.2  ^{+0.7}_{-0.3} $ 
\\
Charge flip 
& $0.3 \pm 0.1  ^{+0.3}_{-0.1} $ 
& $0 \pm 0 ^{+0.1}_{-0.0} $ 
& $0.3 \pm 0.1  ^{+0.2}_{-0.1} $ 
\\
Real
& $0.8 \pm 0  ^{+0.3}_{-0.6} $ 
& $1.0 \pm 0  ^{+0.4}_{-0.6} $
& $2.3 \pm 0  ^{+0.8}_{-1.9} $
\\
\hline
Total
& $1.4 \pm 0.3  ^{+0.4}_{-0.6} $ 
& $1.7 \pm 0.3  \pm 0.7 $
& $3.1 \pm 0.2  ^{+1.1}_{-1.9} $
\\ 
Data 
& 1 
& 2 
& 2 \\ \hline
$tt_{LL}$ 
& $0.2 \pm 0 \pm 0.1$
& $0.2 \pm 0 \pm 0.1$
& $0.5 \pm 0 \pm 0.3$
\\
$tt_{LR}$ 
& $0.02 \pm 0 \pm 0.01$
& $0.001\pm 0 ^{+0.01}_{-0.001}$
& $0.02 \pm 0 \pm 0.02$
\\
$tt_{RR}$ 
& $0.5 \pm 0 \pm 0.3$
& $0.1 \pm 0 \pm 0.2$
& $0.8 \pm 0 \pm 0.3$
\\
$b'$ 450 GeV
& $1.8 \pm 0 \pm 0.3$
& $2.1 \pm 0 \pm 0.3$
& $4.3 \pm 0 \pm 0.5$
\\
\hline\hline
\end{tabular}
}
\caption{Predicted number of SM background events and observed data  with two
  negatively-charged leptons, at least two jets, \MET\ $>$ 40~GeV and $H_{\textrm T}>350$ GeV. Uncertainties are
  statistical followed by systematic.   The expected contribution from
  same-sign top-quark pairs is shown (assuming 
  $C/\Lambda^2 = 1/{\textrm{TeV}}^2$, see Eq.~\ref{lagrange}) as well as 
  from a 450 GeV $b'$.}
\label{tab:sig_neg}
\end{table}

\begin{table}
\center
{\small
\begin{tabular}{lrrr}
\hline\hline
 & $e^+e^+$  & $\mu^+\mu^+$  & $e^+\mu^+$ \\ \hline
Fake 
& $0.8 \pm 0.6  ^{+0.2}_{-0.4} $ 
& $1.0 \pm 0.3  ^{+0.6}_{-0.4} $  %+0.80
& $3.3 \pm 1.1  ^{+1.6}_{-1.4} $  %+1.80
\\
Charge flip 
& $0.3 \pm 0.1  ^{+0.3}_{-0.1} $ 
& $0    \pm 0   ^{+0.1}_{-0.0}$ 
& $0.4 \pm 0.1  ^{+0.3}_{-0.1} $ 
\\
Real 
& $1.9 \pm 0  ^{+0.7}_{-1.5} $ 
& $1.6 \pm 0  ^{+0.7}_{-0.9} $
& $4.4 \pm 0  ^{+1.3}_{-3.1} $
\\
\hline
Total
& $3.0 \pm 0.6  ^{+0.8}_{-1.5} $ 
& $2.6 \pm 0.3  ^{+0.9}_{-1.1} $
& $8.1 \pm 1.1  ^{+2.2}_{-3.4} $ %+1.97
\\ 
Data 
& 2 
& 1 
& 10 
\\ \hline 
$tt_{LL}$ %(1.77 pb)
& $30.1 \pm 0 \pm 5.0$
& $30.4 \pm 0 \pm 4.8$
& $64.2 \pm 0 \pm 10.3$
\\
$tt_{LR}$ %(0.22 pb)
& $3.8 \pm 0 \pm 0.6$
& $4.2 \pm 0 \pm 0.7$
& $8.3 \pm 0 \pm 1.3$
\\
$tt_{RR}$ %(1.77 pb)
& $35.5 \pm 0 \pm 6.0$
& $29.5 \pm 0 \pm 4.6$
& $65.7 \pm 0 \pm 10.4$
\\
$b'$ 450 GeV
& $1.8 \pm 0 \pm 0.3$
& $2.7 \pm 0 \pm 0.4$
& $5.0 \pm 0 \pm 0.7$\\
\hline\hline
\end{tabular}
}
\caption{Predicted number of SM background events and observed data  with two
  positively-charged leptons, at least two jets, \MET\ $>$ 40~GeV and $H_{\textrm T}>350$ GeV. Uncertainties are
  statistical followed by systematic. The expected contribution from
  same-sign top-quark pairs is shown (assuming $C/\Lambda^2 = 1/{\textrm{TeV}}^2$, see Eq.~\ref{lagrange}) as well as 
  from a 450 GeV $b'$.}
\label{tab:sig_pos}
\end{table}

\begin{figure}
  \includegraphics[width=0.46\textwidth]{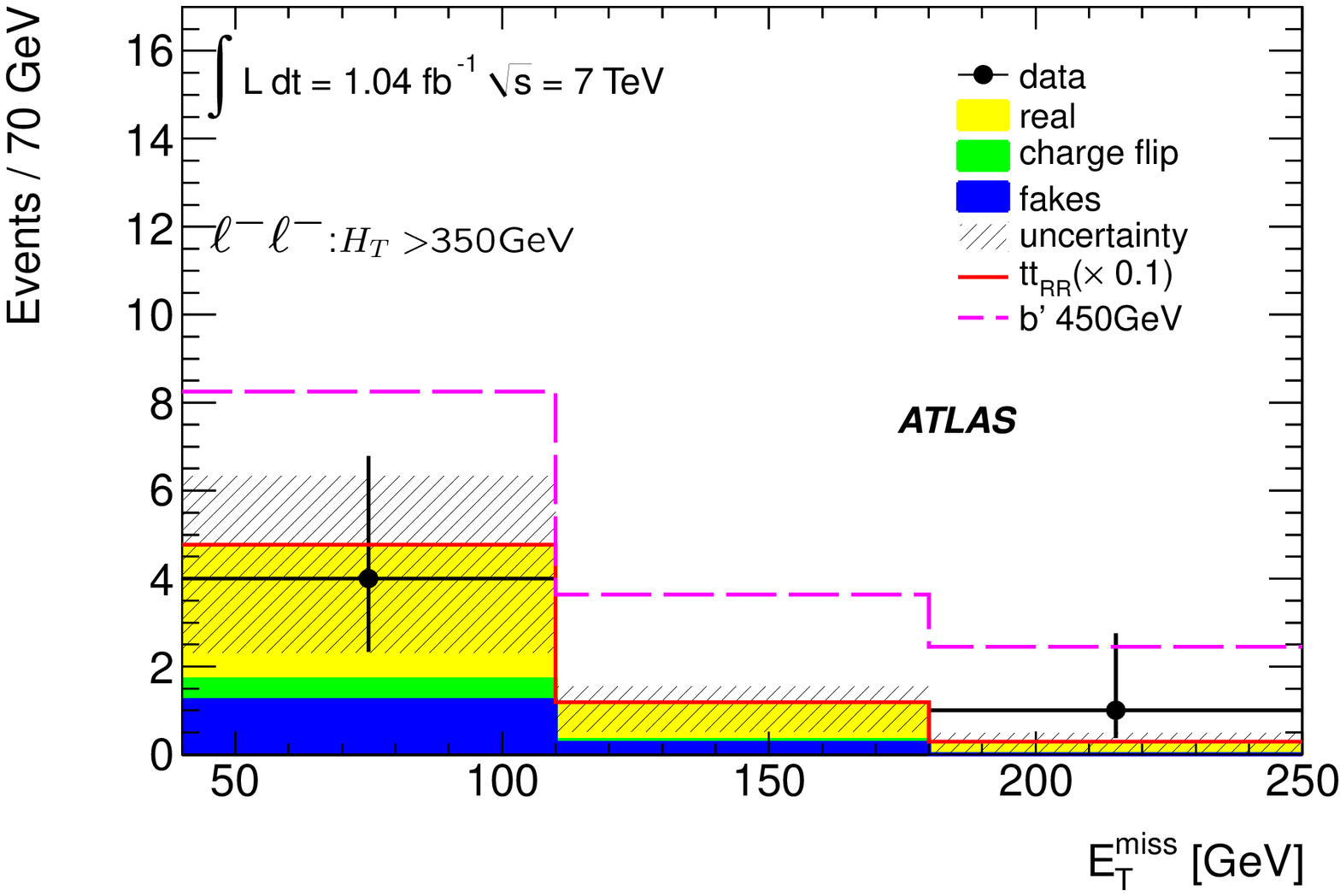}
  \qquad
  \includegraphics[width=0.46\textwidth]{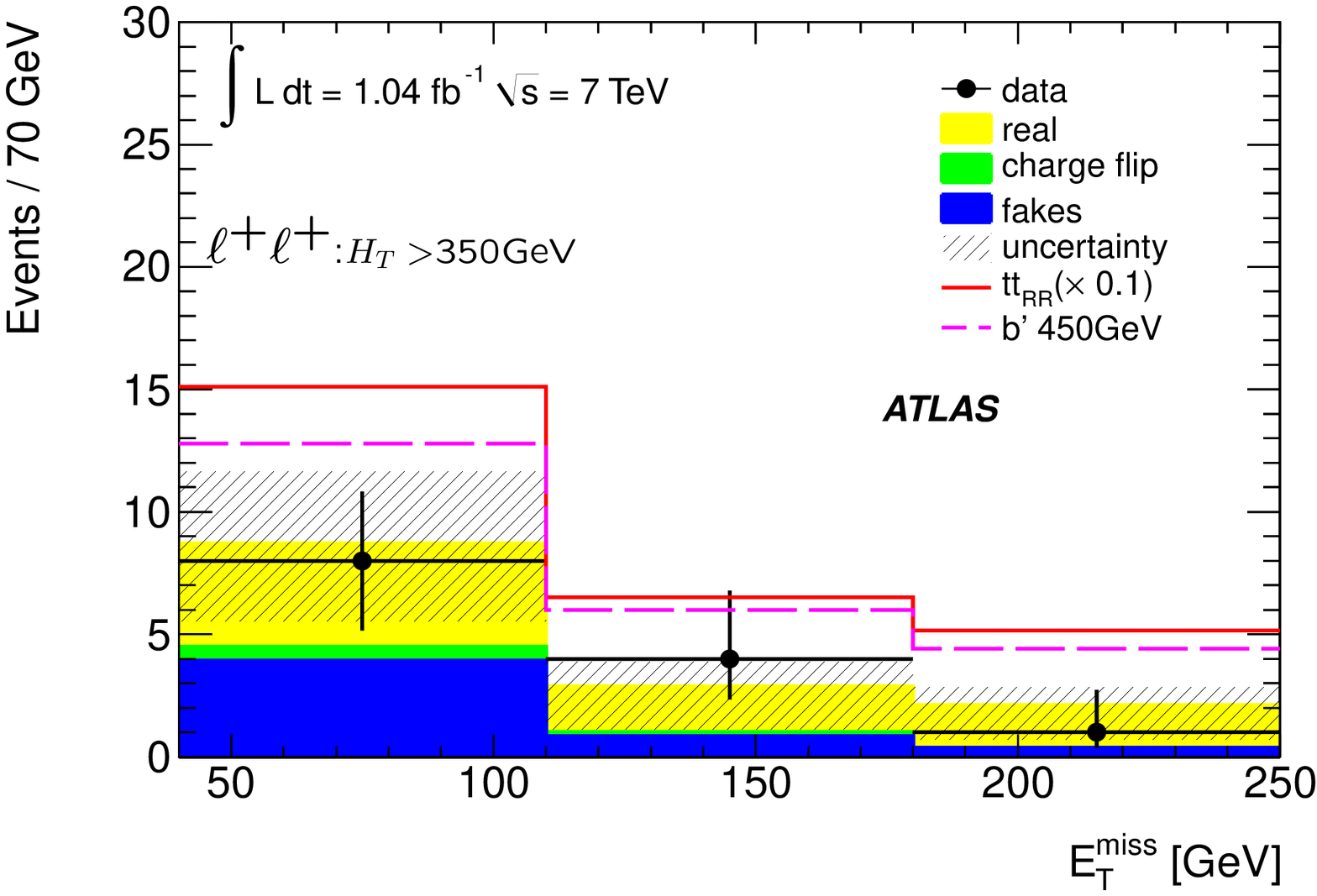}
  \caption{ \MET\ distribution: comparison of observed data and expected SM backgrounds for
    events with a pair of same-sign leptons, at least two reconstructed
    jets, \MET\ $>$ 40~GeV and \HT\ $>$ 350~GeV. Left are negatively-charged lepton pairs, right are
    positively-charged lepton pairs. Uncertainties (hatched) are systematic and statistical. The last bin
    includes overflow events. $tt_{RR}$ (scaled by $0.1$ and assuming
    $C/\Lambda^2=1/{\textrm{TeV}}^2$, see Eq.~\ref{lagrange})  and
    $b'$ signals include both signal and background.}
  \label{fig:sig_ss}
\end{figure}

\begin{figure}
  \includegraphics[width=0.46\linewidth]{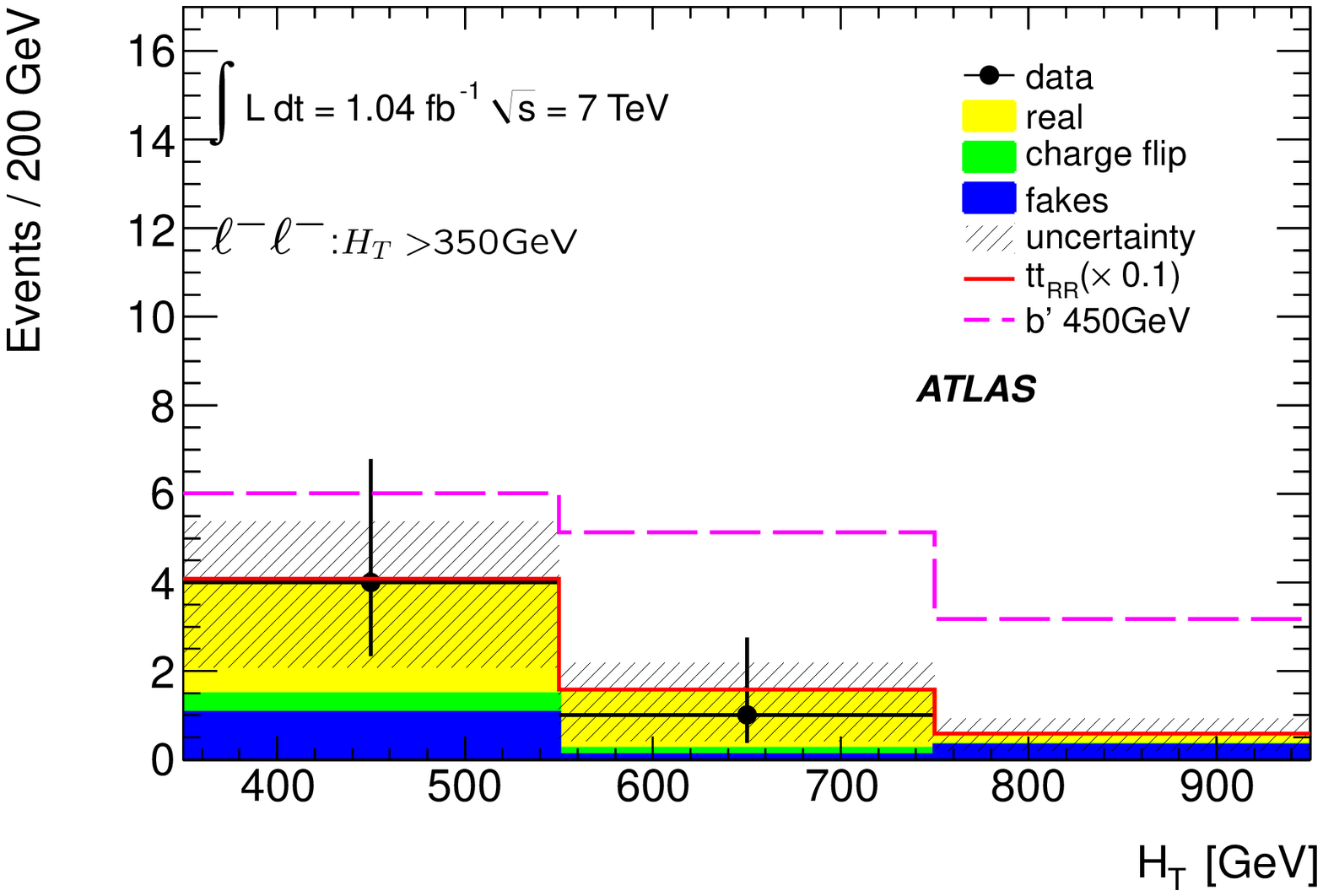}
  \qquad
  \includegraphics[width=0.46\linewidth]{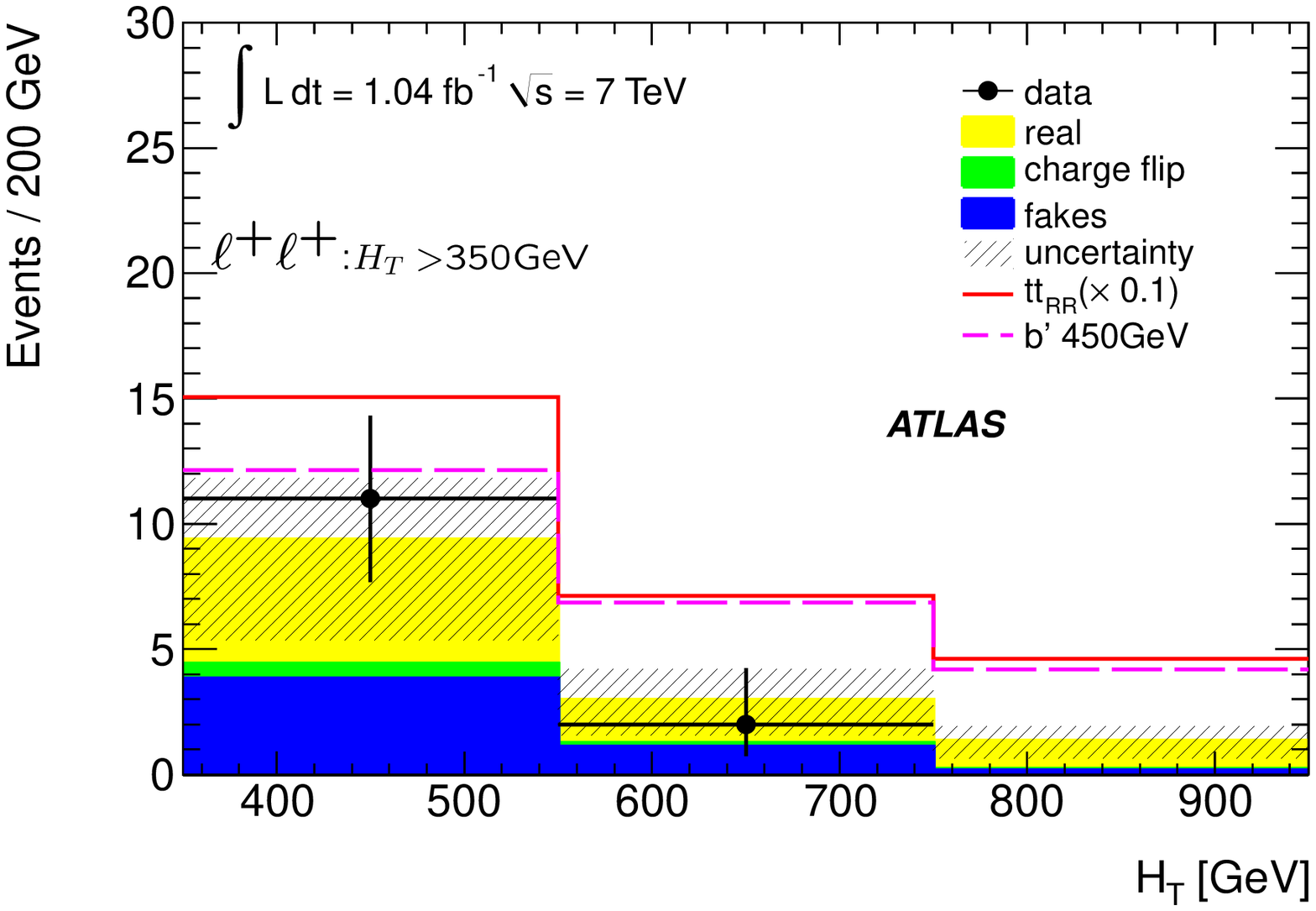}
\caption{ \HT\ distribution: comparison of observed data and expected SM backgrounds for
    events with a pair of same-sign leptons, at least two reconstructed
    jets, \MET\ $>$ 40~GeV and \HT\ $>$ 350~GeV. Left are negatively-charged lepton pairs, right are
    positively-charged lepton pairs. Uncertainties (hatched) are
    systematic and statistical. The last bin
    includes overflow events. $tt_{RR}$ (scaled by $0.1$ and assuming
  $C/\Lambda^2=1/{\textrm{TeV}}^2$, see Eq.~\ref{lagrange}) and $b'$ signals include both signal and background. }
  \label{fig:sig_ss2}
\end{figure}

In the signal region tuned for same-sign top quarks due to low-mass $Z'$ exchange, the expected and observed yields are
shown for events with two negatively (positively) charged leptons in
table~\ref{tab:sig2_neg} (table~\ref{tab:sig2_pos}). The distribution of \MET\
and \HT\ are shown in figures~\ref{fig:sig2_ss_met} and~\ref{fig:sig2_ss_ht}.

\begin{table}
\center
{\small
\begin{tabular}{lrrr}
\hline\hline
 & $e^-e^-$  & $\mu^-\mu^-$  & $e^-\mu^-$ 
\\ \hline
Fake
& $0.2 \pm 0.4  \pm 0.1 $ 
& $0.8 \pm 0.4  ^{+0.4}_{-0.3} $  
& $0.4 \pm 0.3  ^{+1.0}_{-0.1} $  
\\
Charge flip 
& $0.7 \pm 0.1  ^{+0.7}_{-0.2} $ 
& $0.2 \pm 0  \pm 0.2$ 
& $0.5 \pm 0.1  ^{+0.5}_{-0.1} $ 
\\
Real 
& $1.5 \pm 0  ^{+0.6}_{-1.0} $ 
& $1.4 \pm 0  ^{+0.5}_{-0.8} $
& $2.9 \pm 0  ^{+0.8}_{-1.8} $
\\ 
\hline
Total
& $2.4 \pm 0.4  ^{+0.9}_{-1.1} $ 
& $2.4 \pm 0.4  ^{+0.6}_{-0.8} $
& $3.9 \pm 0.3  ^{+1.4}_{-1.8} $ 
\\ 
Data 
& 1 
& 3 
& 1 
\\ \hline
$tt_{RR,m_{Z'}=100{\rm GeV}}$ 
& $0.01 \pm 0 \pm 0.01$
& $0.01 \pm 0 \pm 0.01$
& $0.01 \pm 0 \pm 0.01$
\\
$tt_{RR,m_{Z'}=150{\rm GeV}}$
& $0.02 \pm 0 \pm 0.01$
& $0.02 \pm 0 \pm 0.01$
& $0.03 \pm 0 \pm 0.01$
\\
$tt_{RR,m_{Z'}=200{\rm GeV}}$
& $0.01 \pm 0 \pm 0.01$
& $0.01 \pm 0 \pm 0.01$
& $0.06 \pm 0 \pm 0.02$
\\
\hline\hline
\end{tabular}
}
\caption{Predicted number of SM background events and observed data with two
  negatively-charged leptons, at least two jets, \MET\ $>$ 40~GeV,
  $H_{\textrm T}>150$ GeV and $m_{\ell\ell}>100$ GeV. Uncertainties are
  statistical followed by systematic. The expected contribution from
  same-sign top-quark pairs with low-mass $Z'$ 
exchange is shown, using a fixed value $C/\Lambda^2 = -1~\mathrm{TeV}^{-2}$, 
where $\Lambda=m_{Z'}$ in each case, see Eq.~\ref{lagrange}.}
\label{tab:sig2_neg}
\end{table}

\begin{table}
\center
{\small
\begin{tabular}{lrrr}
\hline\hline
 & $e^+e^+$  & $\mu^+\mu^+$  & $e^+\mu^+$ \\ \hline
Fake 
& $0.5 \pm 0.4  ^{+0.1}_{-0.2} $ 
& $1.6 \pm 0.5  ^{+1.2}_{-1.3} $
& $3.1 \pm 1.0  ^{+2.2}_{-1.5} $
\\
Charge flip 
& $0.6 \pm 0.1  ^{+0.6}_{-0.2} $ 
& $0    \pm 0   ^{+0.1}_{-0.0}$ 
& $0.9 \pm 0.1  ^{+0.6}_{-0.2} $ 
\\
Real 
& $1.9 \pm 0  ^{+0.7}_{-1.1} $ 
& $2.1 \pm 0  ^{+0.8}_{-0.7} $
& $5.6 \pm 0  ^{+1.6}_{-3.5} $
\\
\hline
Total
& $3.0 \pm 0.4  ^{+0.9}_{-1.2} $ 
& $3.7 \pm 0.5  \pm 1.5 $
& $9.6 \pm 1.0  ^{+2.8}_{-3.8} $
\\ 
Data 
& 3 
& 4 
& 8 
\\ \hline 

$tt_{RR,m_{Z'}=100{\rm GeV}}$
& $0.24 \pm 0 \pm 0.04$
& $0.36 \pm 0 \pm 0.06$
& $0.6 \pm 0 \pm 0.1$
\\
$tt_{RR,m_{Z'}=150{\rm GeV}}$
& $0.6 \pm 0 \pm 0.1$
& $0.9 \pm 0 \pm 0.2$
& $1.5 \pm 0 \pm 0.3$
\\
$tt_{RR,m_{Z'}=200{\rm GeV}}$
& $1.1 \pm 0 \pm 0.2$
& $1.7 \pm 0 \pm 0.3$
& $3.2 \pm 0 \pm 0.6$

\\
\hline\hline
\end{tabular}
}
\caption{Predicted number of SM background events and observed data with two
  positively-charged leptons, at least two jets, \MET\ $>$ 40~GeV,
  $H_{\textrm T}>150$ GeV and $m_{\ell\ell}>100$ GeV. Uncertainties are
  statistical followed by systematic. The expected contribution
   from same-sign top-quark pairs with low-mass $Z'$ exchange is shown,
   using a fixed value of the coupling $C/\Lambda^2= - 1~\mathrm{TeV}^{-2}$, where
   $\Lambda=m_{Z'}$ in each case, see Eq.~\ref{lagrange}.}
\label{tab:sig2_pos}
\end{table}

\begin{figure}
  \includegraphics[width=0.46\linewidth]{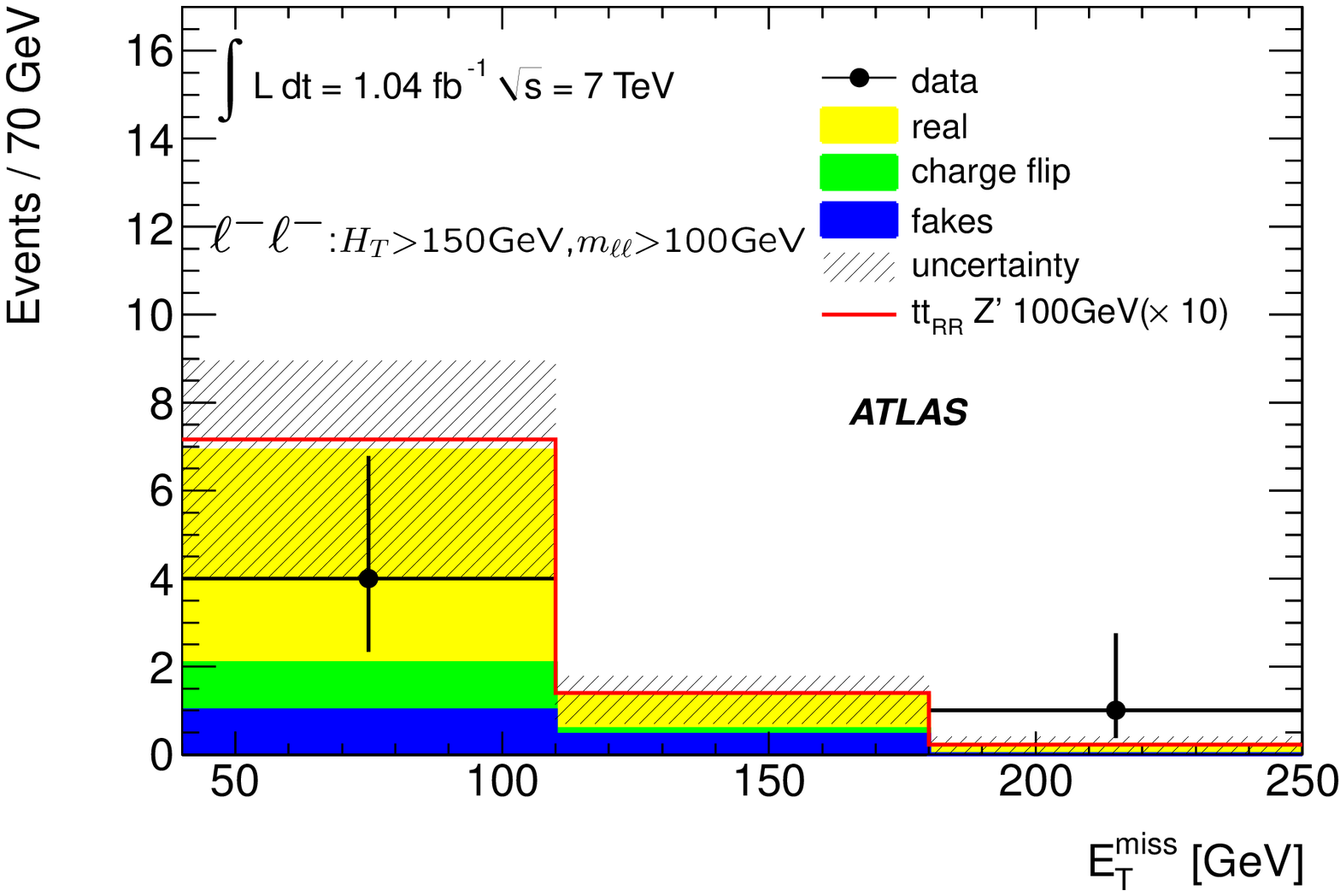}
  \qquad
  \includegraphics[width=0.46\linewidth]{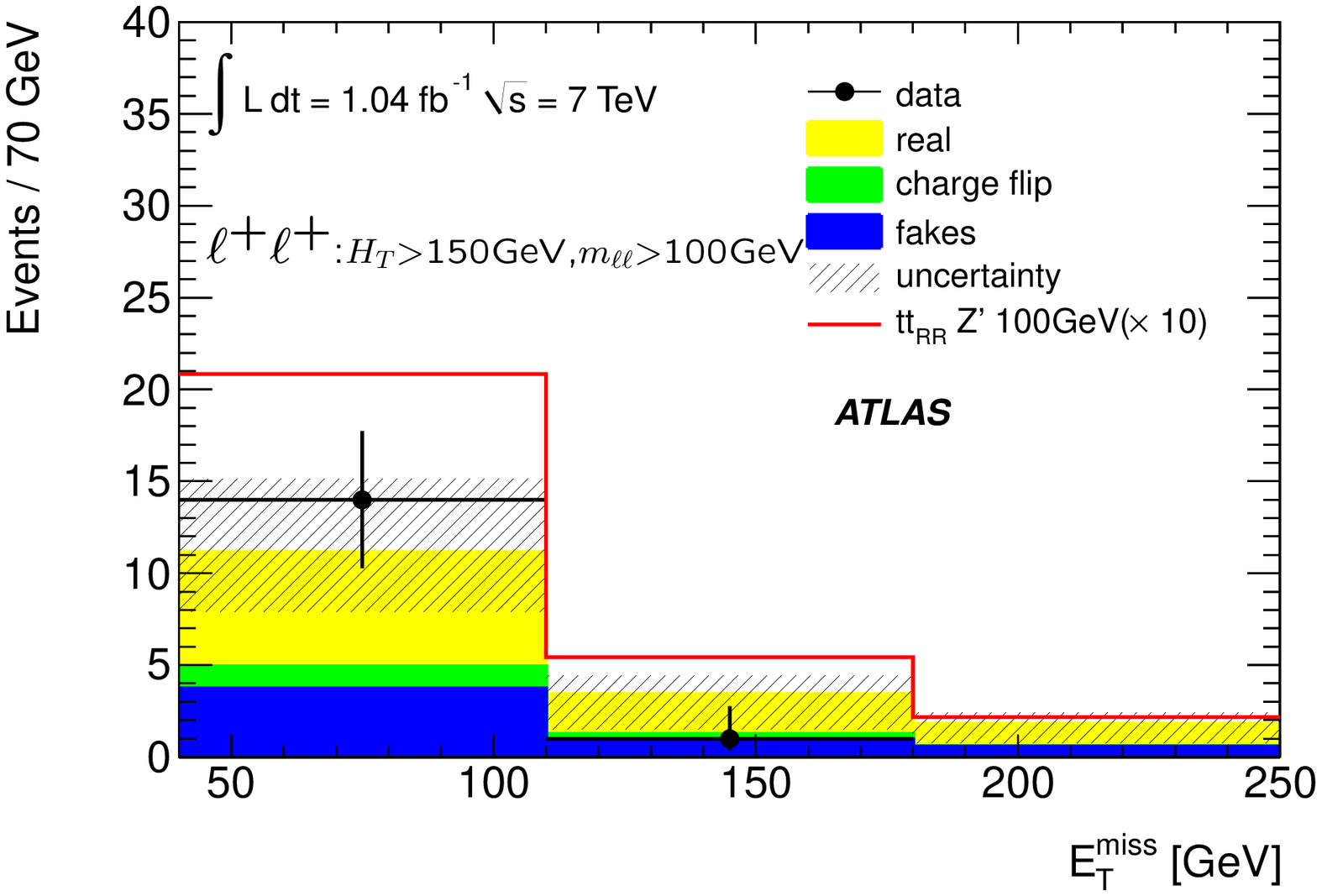}
  \caption{ \MET\ distribution: comparison of observed data and expected SM backgrounds for
    events with a pair of same-sign  leptons, at least two reconstructed
    jets, \MET\ $>$ 40~GeV, \HT\ $>$ 150~GeV and
    $m_{\ell\ell}>100$~GeV. Left are negatively-charged lepton pairs, right are
    positively-charged lepton pairs. Uncertainties (hatched) are systematic and statistical. The last bin
    includes overflow events. $tt_{RR}$ from $Z'$ 100 GeV ($\times
    10$) signal histogram includes both signal and background.}
  \label{fig:sig2_ss_met}
\end{figure}

\begin{figure}
  \includegraphics[width=0.46\linewidth]{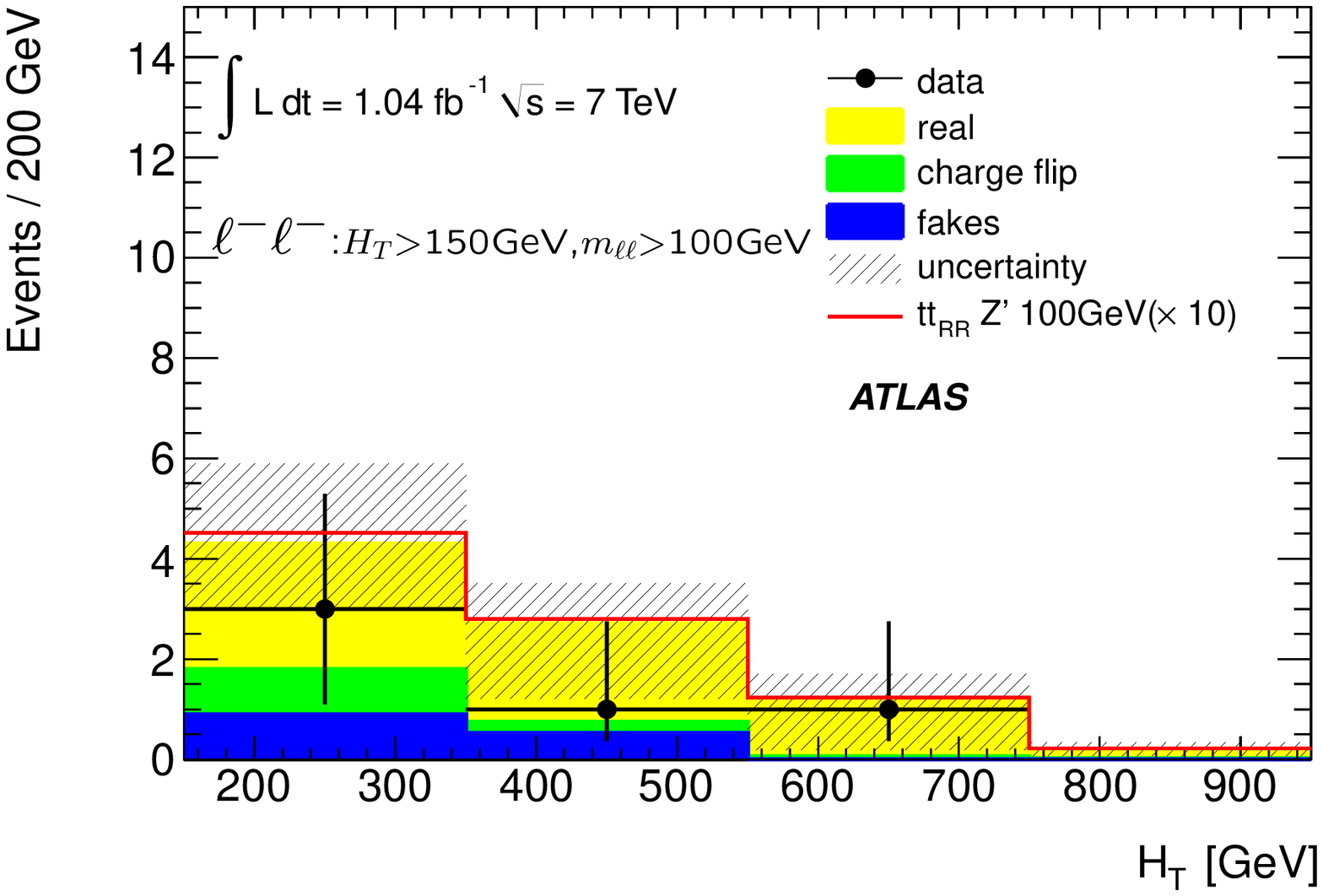}
  \qquad
  \includegraphics[width=0.46\linewidth]{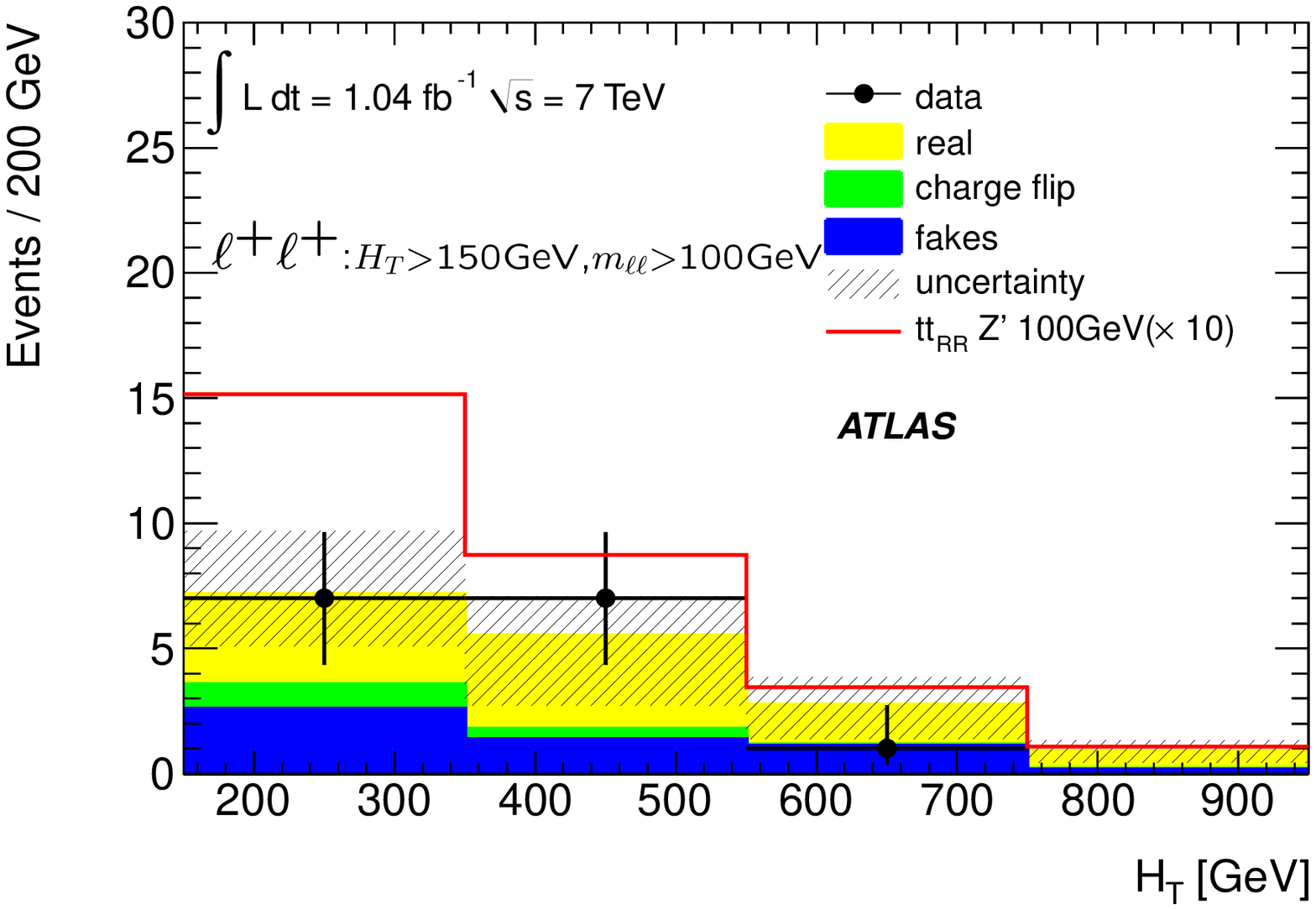}
  \caption{ \HT\  distributions: comparison of observed data and expected SM backgrounds for
    events with a pair of same-sign  leptons, at least two reconstructed
    jets, \MET\ $>$ 40~GeV, \HT\ $>$ 150~GeV and
    $m_{\ell\ell}>100$~GeV. Left are negatively-charged lepton pairs, right are
    positively-charged lepton pairs. Uncertainties (hatched) are systematic and statistical. The last bin
    includes overflow events. $tt_{RR}$ from $Z'$ 100 GeV ($\times
    10$) signal histogram includes both signal and background.}
  \label{fig:sig2_ss_ht}
\end{figure}

\clearpage
\section{Interpretation of the Results}
Since the data are consistent with the Standard Model expectations,
the analysis sets limits on the production of two processes producing same-sign dilepton
signals from new physics sources. For each model, upper limits at 95\%
confidence level on the cross sections of the hypothetical processes are derived
using the \textit{CL}s method~\cite{cls1,cls2}. In both cases, we use a
single-bin counting experiment, fitting the data to extract the most
likely signal cross section. Systematic uncertainties are included as variations in
the expected signal and background yields, which are fluctuated in the
ensembles used to generate the \textit{CL}s distributions. 

\subsection{Same-sign top-quark production}

We calculate upper limits on the cross section of same-sign top-quark pair
production using only the positively-charged lepton pairs. Modeling $tt$ production in terms of effective four-fermion operators, the
expected $95\%$ confidence level limits on $\sigma(pp\rightarrow tt)$ are shown in
table~\ref{tab:sig_tt} for the three possible chirality combinations of the $tt$ pair, which
influence the efficiency primarily through the lepton transverse
momentum. In table~\ref{tab:sig2_tt} the limits are given for a model
with a light flavour-changing $Z'$ boson with right-handed couplings,
and three values of its mass.  These limits supersede those on this process previously reported by ATLAS~\cite{atlasssmuon}. The limits reported here are more stringent than those of Ref~\cite{atlasssmuon} due to the use of both electrons and muons, and an event selection optimized for same-sign top-quark pair production including a jet multiplicity requirement. Ref~\cite{atlasssmuon} used a more inclusive selection examining a range of new physics models.

\begin{table}
\center
\begin{tabular}{lcccc}
\hline\hline
Chirality  & Median expected & 68\% range & Observed &Observed
 \\ 
config.  &  limit, $\sigma$  &  limit, $\sigma$&  limit, $\sigma$ &  limit, $C$ \\ \hline
LL & $\sigma<1.8$ pb & 1.1-3.2 pb & $\sigma<1.7 $ pb & $C_{LL}/\Lambda^{2}<0.35$ $\rm TeV^{-2}$\\ 
LR & $\sigma<1.7$ pb& 1.0-3.0 pb  &$\sigma<1.7$ pb &  $C_{RL}/\Lambda^{2},C'_{LR}/\Lambda^{2}<0.98$ $\rm TeV^{-2}$\\ 
RR & $\sigma<1.7$ pb & 1.0-3.0 pb &$\sigma<1.7$ pb &  $C_{RR}/\Lambda^{2}<0.35$ $\rm TeV^{-2}$\\ 
\hline\hline
\end{tabular}
\caption{Expected and observed upper limits on same-sign top-quark cross
  section  at 95\% confidence level. The uncertainties for the expected limits
  describe a range which includes 68\% of pseudo-experiments drawn
  from the background-only hypothesis. A dileptonic decay branching
  ratio of 10.6\% has been taken into account, so that the limits
  are directly on $\sigma(pp\rightarrow tt)$. The observed limit on the coefficients $C_{LL},C_{LR}=C'_{RL},C_{RR}$ of the effective operator is also indicated.}
\label{tab:sig_tt}
\end{table}

\begin{table}
\center
\begin{tabular}{lccc}
\hline\hline
$Z'$ mass  & Median expected & 68\% range & Observed 
 \\ 
  &  limit, $\sigma$ &  limit, $\sigma$  &  limit, $\sigma$  \\ \hline
100 GeV & $\sigma<2.1$ pb & 1.4-3.3 pb & $\sigma<2.0 $ pb \\ 
150 GeV & $\sigma<1.7$ pb & 1.0-2.8 pb& $\sigma<1.6 $ pb \\ 
200 GeV & $\sigma<1.5$ pb & 0.9-2.3 pb& $\sigma<1.4 $ pb \\ 
\hline\hline
\end{tabular}
\caption{Expected and observed upper limits on same-sign top-quark production from low mass $Z'$ cross
  sections at 95\% confidence level. The uncertainties for the expected limits
  describe a range which includes 68\% of pseudo-experiments drawn
  from the background-only hypothesis. A dileptonic decay branching ratio of 10.6\% has been taken into account.}
\label{tab:sig2_tt}
\end{table}

The cross-section limits in table \ref{tab:sig_tt} can be directly translated into limits on
coefficients of effective operators corresponding to each pair of
chiralities. There are five independent dimension-six four-fermion operators mediating $uu \to tt$~\cite{sstops2}, with four possible structures. The resulting Lagrangian relevant for $tt$ production reads
\begin{eqnarray}
\mathcal{L}_{4F} & = &
 {\oh} \frac{C_{LL}}{\Lambda^2} (\bar u_L \gM t_L) (\bar u_L \gm t_L)
 + {\oh} \frac{C_{RR}}{\Lambda^2} (\bar u_R \gM t_R) (\bar u_R \gm t_R) \notag \\
 & &  - {\oh} \frac{C_{LR}}{\Lambda^2} (\bar u_L \gM t_L) (\bar u_R \gm t_R)
 -{\oh}  \frac{C'_{LR}}{\Lambda^2} (\bar u_{La} \gM t_{Lb}) (\bar
 u_{Rb} \gm t_{Ra}) + \text{h.c.} \,,
\label{lagrange}
\end{eqnarray}
where the subindices $a,b$ in the last term indicate the colour
contractions. The limits on operator coefficients are the most relevant for a comparison between
different experiments, because the production cross sections 
themselves depend on the type of particles being collided and their centre-of-mass
energy. These limits improve on those reported by the CMS
Collaboration, $C_{RR}/\Lambda^2 <
2.7~\text{TeV}^{-2}$~\cite{Chatrchyan:2011dk} using 35 \ipb\ of 2010 data.

Constraints can also be placed on generic classes of models with new
particles mediating same-sign top-quark production at the tree level. These new particles can be classified according to their quantum numbers under the SM group
$\text{SU}(3)_C \times \text{SU}(2)_L \times
\text{U}(1)_Y$. In table~\ref{tab:lim-rest}
limits are given for different types of particles, using the notation of
Ref.~\cite{sseffop}; the quantum numbers $(C,I)_Y$ refer to the
transformation properties under the SM group, where $C$ is the colour
(octet, sextet, triplet or singlet); $I$ the weak isospin (triplet, doublet
or singlet) and $Y$ is the hypercharge. The couplings labeled as $g_{13}$
involve a flavour change between the up and top quark, whereas
$g_{11}$, $g_{33}$ are flavour-diagonal couplings of the new particle
to the up and top quarks, respectively; for $\bmu$ and $\gmu$ vector
bosons $|g_{13}| \equiv \left(|g_{13}^q|^2+|g_{13}^u|^2
\right)^{1/2}$, where $g_{13}^q$ is left-handed and $g_{13}^u$ is right-handed.
In particular, limits are placed on new colour-sextet scalars and
vector bosons produced in the $s$-channel~\cite{Berger:2010fy}, as well as on heavy flavour-changing scalars and vector bosons.
The last column gives the mass limits for unit couplings. Notice that for larger couplings the mass limits are more stringent, and conversely for smaller couplings the mass limits are looser.

\begin{table}[htb]
\begin{center}
\begin{tabular}{ccccc} \hline\hline
Label & Spin & Quantum numbers & Limit & Mass Limit \\[1mm]
\hline \\[-4mm]
$\bmu$ & 1 & $(1,1)_0$ & $|g_{13}|/\Lambda < 0.57$ TeV$^{-1}$ & 1.7 TeV \\[1mm]
$\wmu$ & 1 & $(1,3)_0$ & $|g_{13}|/\Lambda < 0.57$ TeV$^{-1}$ & 1.7 TeV \\[1mm]
$\gmu$ & 1 & $(8,1)_0$ & $|g_{13}|/\Lambda < 0.99$ TeV$^{-1}$ & 1.0 TeV \\[1mm]
$\hmu$ & 1 & $(8,3)_0$ & $|g_{13}|/\Lambda < 0.99$ TeV$^{-1}$ & 1.0 TeV \\[1mm]
$\qmu$ & 1 & $(3,2)_{-\frac{5}{6}}$ & $|g_{11} g_{33}|/\Lambda^2 < 0.34$ TeV$^{-2}$ & 1.7 TeV \\[1mm]
$\ymu$ & 1 & $(\bar 6,2)_{-\frac{5}{6}}$ & $|g_{11} g_{33}|/\Lambda^2 < 0.63$ TeV$^{-2}$  & 1.3 TeV \\[1mm]
$\phi$ & 0 & $(1,2)_{-\frac{1}{2}}$ & $|g_{13}^u g_{31}^u|/\Lambda^2 < 0.92$ TeV$^{-2}$ & 1.1 TeV \\[1mm]
$\Phi$ & 0 & $(8,2)_{-\frac{1}{2}}$ & $|g_{13}^u g_{31}^u|/\Lambda^2 < 1.8$ TeV$^{-2}$ & 0.8 TeV \\[1mm]
$\Of$  & 0 & $(\bar 6,1)_{-\frac{4}{3}}$ & $|g_{11} g_{33}|/\Lambda^2 <0.33$ TeV$^{-2}$  & 1.8 TeV \\[1mm]
$\So$ & 0 & $(\bar 6,3)_{-\frac{1}{3}}$ & $|g_{11} g_{33}|/\Lambda^2 < 0.16$ TeV$^{-2}$ & 2.5 TeV \\
\hline\hline
\end{tabular}
\end{center}
\caption{Lower (upper) limits at 95\% confidence level on the masses (couplings) for generic heavy vector bosons and
  scalars which mediate the production of same-sign top-quark pairs. A theoretical uncertainty due to variations of the
  $Q^2$ scale gives a 5\% uncertainty on the limits of the couplings
  $g$. Quantum numbers are defined in the text.}
\label{tab:lim-rest}
\end{table}

Limits for neutral colour singlets ($\bmu$) are of interest because
the exchange of a \mbox{$t$-channel} $Z'$ boson, corresponding to $\bmu$ in
table~\ref{tab:lim-rest}, has been proposed as a possible mechanism
which could increase the value of the forward-backward production
asymmetry $A_{\textrm{FB}}$ in $t\bar{t}$ production at the Tevatron.
CDF and D0 have reported measurements~\cite{Aaltonen:2011kc,Collaboration:2011rq} of this asymmetry which lie above the Standard Model expectation. In the simplest realisation of this idea,
the $Z'$ boson is real and leads to same-sign top-quark pair production~\cite{Jung:2009jz,Berger:2011ua}.
For a given mass and coupling of the new $Z'$ boson, its contribution to the
$A_{\textrm{FB}}$ at the Tevatron and the $tt$ cross section at the LHC are related. Figure~\ref{fig:lim-AFB} shows the contributions to the inclusive 
(left) and high-mass (right) $A_\textrm{FB}$ from the exchange of a $Z'$ boson 
with right-handed couplings, versus the same-sign top-quark pair cross sections at the
LHC. They have been obtained following the method outlined in 
Ref.~\cite{sseffop2}.
\begin{figure}[htb]
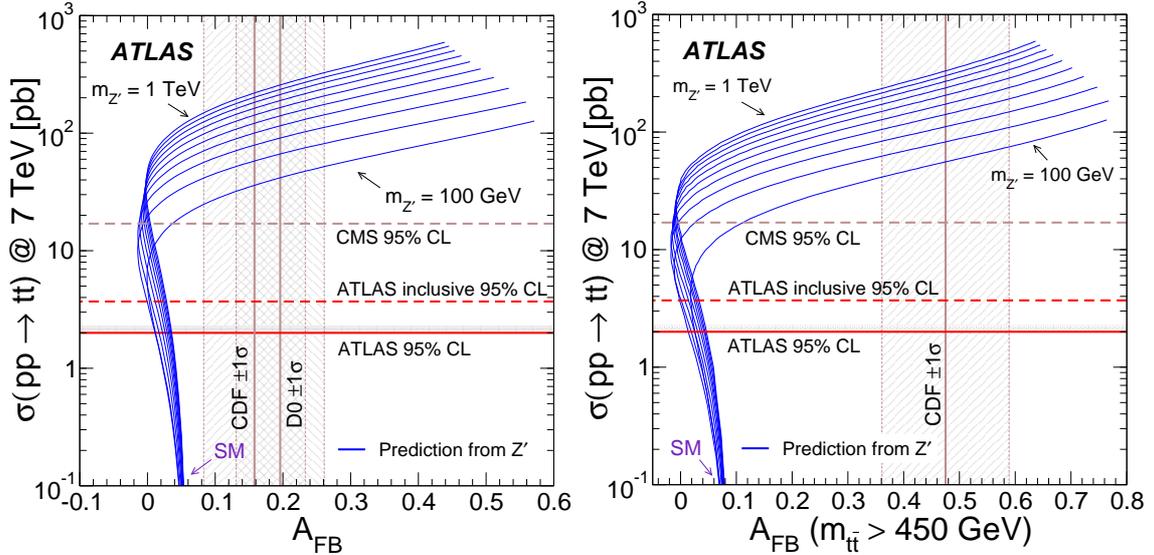

\begin{center}
\includegraphics[width=0.49\linewidth,clip=]{afb.eps}
\includegraphics[width=0.49\linewidth,clip=]{afbH.eps}
\end{center}
\caption{Allowed regions for the new physics contributions to the inclusive 
(left) and high-mass (right) $A_{\textrm{FB}}$ at Tevatron, and the $tt$ cross 
section at LHC.  Limits from this analysis are the solid 
horizontal lines.  The  measurements of $A_{\textrm{FB}}$ from CDF and
D0 are shown as vertical lines with bands representing the uncertainties. }
\label{fig:lim-AFB}
\end{figure}

 For each curve in figure~\ref{fig:lim-AFB},  corresponding to 
a different $Z'$ boson mass ranging between 100 GeV and 1 TeV from bottom to
top, the lower end (out of scale) corresponds to vanishing 
coupling $g_{13}=0$.  The shape of the curves is due to  the interference of the $Z'$
with the $t\bar{t}$ production amplitude, which gives a negative
contribution to the forward-backward production asymmetry, while the 
quadratic $Z'$ contribution increases it. As the coupling is increased 
from zero, the contribution to the forward-backward asymmetry is first
negative and then becomes positive at larger couplings. 

The solid horizontal line represents the 95\% CL limit on same-sign top-quark 
production obtained from this analysis, taken as the most conservative one in 
tables~\ref{tab:sig_tt} and~\ref{tab:sig2_tt}. For comparison, limits from previous analyses~\cite{Chatrchyan:2011dk,atlasssmuon} are also shown. 
Previous ATLAS limits~\cite{atlasssmuon} already exclude the possibility of any positive 
contributions to either the inclusive or the high-mass $A_{\textrm{FB}}$ in $t\bar{t}$ 
production in minimal models with a single, real $Z'$ boson. Still, these 
constraints can be evaded in non-minimal models~\cite{Jung:2011zv}
which involve more than one $Z'$ boson so as to partially cancel their contribution to same-sign top quark
production. The tighter constraints reported here narrow the parameter space 
for such cancellations. 

\subsection{Heavy down-type quarks}

The second process investigated is pair production of chiral
down-type charge $-1/3$ quarks $b'$ with decay $b' \bar b' \to W^-t
W^+ \bar t$. The branching ratio for $b' \to Wt$ is assumed to be
unity for fourth-generation quarks $b'$ if $m_{b'}
> m_{t'}+m_W$ or for other heavy quark models such as exotic isodoublets $(T_B)_{L,R}$ coupling
predominantly to the third generation~\cite{t53}. The cross section for strong pair
production is the same in both cases. % The signal was modeled using fourth-generation quarks and generated using {\sc pythia}. 
The observed and expected cross-section limits for $b' \bar b'$
production are shown in figure~\ref{fig:bprimelimit}. The intersection
of the limit with the theoretical cross-section calculation at next-to-next-to-leading order in QCD yields a lower bound of 450 \gev\ on the new quark mass.

\begin{figure}[htp]
  \centering
  \includegraphics[width=0.6\textwidth]{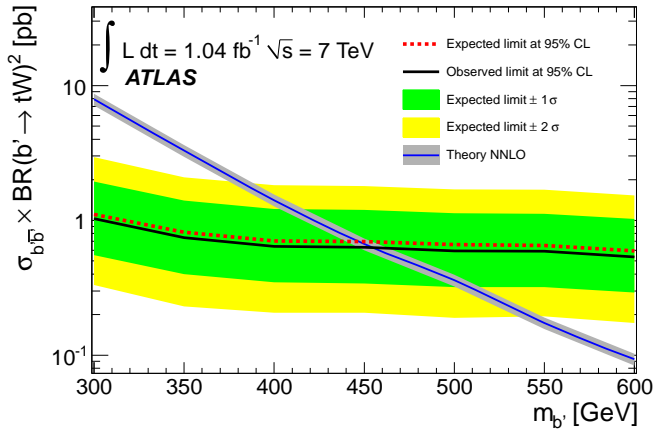} 
  \caption{$95\%$ confidence level exclusion limits on cross section times branching
    ratio for $b' \bar b'$ production with decay $b'\rightarrow tW$.}
  \label{fig:bprimelimit}
\end{figure}

\section{Conclusion}
 A search is presented for anomalous same-sign top-quark production
 and heavy fourth-generation down-type quark production in events with two isolated leptons
  ($e$ or $\mu$) having the same electric charge, large missing
  transverse momentum, and at least two jets. The data are
  selected from events collected from $pp$
  collisions at $\sqrt{s}=7$~\tev\  by the ATLAS detector and correspond
  to an integrated luminosity of 1.04~\ifb.  The observed data are
  consistent with expectations from Standard Model processes.
 Upper limits are set at 95\% confidence level on the cross section of new sources of
 same-sign top-quark pair production via a heavy mediator at 1.7 pb for each
 chirality. For light $Z'$ mediators, limits range from 1.4-2.0 pb
 depending on the $Z'$ mass.  These limits are translated into limits
 on coefficients of effective operators which mediate $uu\rightarrow
 tt$ production, and an interpretation is presented for the case of a
 flavor-changing $Z'$ boson, which has been proposed as an explanation
 for the measurement of the  top-quark pair production
 forward-backward asymmetry at the Tevatron.  In addition, limits
 are set on the production of heavy down-type quarks.  A lower limit
 of 450 \gev\ is set at 95\% confidence level on the mass of
 fourth-generation down-type quarks. This is the strongest limit in
 the same-sign lepton channel, complementing a stronger limit in the
 single-lepton channel~\cite{atlasljets}, which has a larger branching
 ratio but significantly higher background levels.

\section{Acknowledgements}

We thank CERN for the very successful operation of the LHC, as well as the
support staff from our institutions without whom ATLAS could not be
operated efficiently.

We acknowledge the support of ANPCyT, Argentina; YerPhI, Armenia; ARC,
Australia; BMWF, Austria; ANAS, Azerbaijan; SSTC, Belarus; CNPq and FAPESP,
Brazil; NSERC, NRC and CFI, Canada; CERN; CONICYT, Chile; CAS, MOST and NSFC,
China; COLCIENCIAS, Colombia; MSMT CR, MPO CR and VSC CR, Czech Republic;
DNRF, DNSRC and Lundbeck Foundation, Denmark; EPLANET and ERC, European Union;
IN2P3-CNRS, CEA-DSM/IRFU, France; GNAS, Georgia; BMBF, DFG, HGF, MPG and AvH
Foundation, Germany; GSRT, Greece; ISF, MINERVA, GIF, DIP and Benoziyo Center,
Israel; INFN, Italy; MEXT and JSPS, Japan; CNRST, Morocco; FOM and NWO,
Netherlands; RCN, Norway; MNiSW, Poland; GRICES and FCT, Portugal; MERYS
(MECTS), Romania; MES of Russia and ROSATOM, Russian Federation; JINR; MSTD,
Serbia; MSSR, Slovakia; ARRS and MVZT, Slovenia; DST/NRF, South Africa;
MICINN, Spain; SRC and Wallenberg Foundation, Sweden; SER, SNSF and Cantons of
Bern and Geneva, Switzerland; NSC, Taiwan; TAEK, Turkey; STFC, the Royal
Society and Leverhulme Trust, United Kingdom; DOE and NSF, United States of
America.

The crucial computing support from all WLCG partners is acknowledged
gratefully, in particular from CERN and the ATLAS Tier-1 facilities at
TRIUMF (Canada), NDGF (Denmark, Norway, Sweden), CC-IN2P3 (France),
KIT/GridKA (Germany), INFN-CNAF (Italy), NL-T1 (Netherlands), PIC (Spain),
ASGC (Taiwan), RAL (UK) and BNL (USA) and in the Tier-2 facilities
worldwide.

\clearpage

% ATLAS Collaboration author list for 05-JAN-2012
% Data extracted on 16-Mar-2012 for paperid 181
%\documentclass[11pt]{article}
%\usepackage{a4wide}\begin{document}
\begin{flushleft}
{\Large The ATLAS Collaboration}

\bigskip

G.~Aad$^{\rm 48}$,
B.~Abbott$^{\rm 110}$,
J.~Abdallah$^{\rm 11}$,
S.~Abdel~Khalek$^{\rm 114}$,
A.A.~Abdelalim$^{\rm 49}$,
A.~Abdesselam$^{\rm 117}$,
O.~Abdinov$^{\rm 10}$,
B.~Abi$^{\rm 111}$,
M.~Abolins$^{\rm 87}$,
O.S.~AbouZeid$^{\rm 157}$
% ...
\bigskip

% ...
$^{10}$ Institute of Physics, Azerbaijan Academy of Sciences, Baku, Azerbaijan\\
$^{11}$ Institut de F\'isica d'Altes Energies and Departament de F\'isica de la Universitat Aut\`onoma  de Barcelona and ICREA, Barcelona, Spain\\
% ...
$^{48}$ Fakult\"{a}t f\"{u}r Mathematik und Physik, Albert-Ludwigs-Universit\"{a}t, Freiburg i.Br., Germany\\
$^{49}$ Section de Physique, Universit\'e de Gen\`eve, Geneva, Switzerland\\
% ...
$^{87}$ Department of Physics and Astronomy, Michigan State University, East Lansing MI, United States of America\\
% ...
$^{110}$ Homer L. Dodge Department of Physics and Astronomy, University of Oklahoma, Norman OK, United States of America\\
$^{111}$ Department of Physics, Oklahoma State University, Stillwater OK, United States of America\\
% ...
$^{114}$ LAL, Univ. Paris-Sud and CNRS/IN2P3, Orsay, France\\
% ...
$^{117}$ Department of Physics, Oxford University, Oxford, United Kingdom\\
% ...
$^{157}$ Department of Physics, University of Toronto, Toronto ON, Canada\\
% ...
\end{flushleft}

%\end{document}
%\end{document}

% ATLAS Collaboration author list for 16-MAR-2011
% Data extracted on 19-May-2011 for paperid 74

\end{document}